\documentclass[aps,pre,reprint, amsmath, amssymb,superscriptaddress]{revtex4-1}
\usepackage{morefloats}
\usepackage{bm}
\newcommand{\beq}{\begin{equation}}
\newcommand{\eeq}{\end{equation}}

\usepackage[retainorgcmds]{IEEEtrantools}
\usepackage{graphicx,tikz,placeins}
\usepackage{mathrsfs}
\usepackage{amsmath,amssymb,amsfonts,physics}
\usepackage{color}
\usepackage{float}
\usepackage{times,txfonts}
\usepackage{nicefrac}
\usepackage[colorlinks=true,linkcolor=blue,urlcolor=blue,citecolor=blue,pdfusetitle]{hyperref}
\usepackage{physics}
\usepackage{soul}

\begin{document}

\title{Splitting of nonequilibrium phase transitions in driven Ising models}
\author{Gustavo A. L. For\~ao}
\affiliation{Universidade de São Paulo,
Instituto de Física,
Rua do Matão, 1371, 05508-090
São Paulo, SP, Brazil}

\author{Fernando S. Filho}
\affiliation{Universidade de São Paulo,
Instituto de Física,
Rua do Matão, 1371, 05508-090
São Paulo, SP, Brazil}

\author{André P. Vieira}
\affiliation{Universidade de São Paulo,
Instituto de Física,
Rua do Matão, 1371, 05508-090
São Paulo, SP, Brazil}

\author{Bart Cleuren}
\affiliation{UHasselt, Faculty of Sciences, Theory Lab, Agoralaan, 3590 Diepenbeek, Belgium}

\author{Daniel M. Busiello}
\affiliation{Max Planck Institute for the Physics of Complex Systems, 01187 Dresden, Germany}
\affiliation{Department of Physics and Astronomy, University of Padova, 35131 Padova, Italy}

\author{Carlos E. Fiore}
\affiliation{Universidade de São Paulo,
Instituto de Física,
Rua do Matão, 1371, 05508-090
São Paulo, SP, Brazil}

\date{\today}

\begin{abstract}
Spontaneous symmetry breaking occurs in various equilibrium and nonequilibrium systems, where phase transitions are typically marked by a single critical point that separates ordered and disordered regimes. We reveal a novel phenomenon in which the interplay between different temperatures and 
driving forces splits the order-disorder transition into two distinct transition points depending on which ordered state initially dominates. Crucially, these two emerging phases have distinct scaling behaviors and thermodynamic properties. To study this, we propose a minimal variant of the Ising model where spins are coupled to two thermal baths and subjected to two opposite driving forces associated to them. Our findings, robust both for all-to-all interactions (where exact solutions are possible) and nearest-neighbor couplings on a square lattice, uncover unique nonequilibrium behaviors and scaling laws for crucial thermodynamic quantities, such as efficiency, dissipation, power and its fluctuations, that are different between the two ordered phases. We also highlight that one of these emerging phases enables heat-engine operations that are less dissipative and show reduced fluctuations. In this setup, the system can also operate near maximum power and efficiency over a wide parameter range. Our results offer new insights into the relevance of phase transitions under nonequilibrium conditions.
\end{abstract}

\maketitle

{\it Introduction}---
Phase transitions and universal scaling are central topics in statistical physics, particularly for systems that display collective behavior. Examples {can be found across diverse} fields, {including} physics, chemistry, biology, and economics. Many of these systems operate out of equilibrium, where non-zero probability currents and {specific} symmetries can lead to unique classes of nonequilibrium phase transitions. Similar to {their equilibrium counterparts}, nonequilibrium phase transitions are often characterized by an order parameter. However, this approach can obscure the irreversible dynamics and its impact on the properties of phase transitions. Increasing attention has been given to entropy production -- a key indicator of system dissipation -- and related thermodynamic quantities as alternative descriptors of nonequilibrium phase transitions, although {many aspects of this topic remain to be understood} \cite{tome2012, noa2019, aguilera2023nonequilibrium, loos2023long}.

Beyond phase transitions and universality classes, systems exhibiting collective dynamics are of considerable interest for their potential to enhance the performance of nonequilibrium heat engines, both in classical \cite{herpich, herpich2, filho2023powerful, mamede2023, Gashu024, PhysRevE.107.044102, liang2023minimal, rolandi} and quantum thermodynamics \cite{PhysRevE.107.044102, campisi2016power, souza2022collective}. A key subset of collective-engine setups {extract} work from {driving} forces, often modeled {as} biases over specific transitions \cite{herpich, herpich2, gatien, gatien2, filho2023powerful, mamede2023}. {These biases arise} in diverse systems, {such as} 
{anomalous mobility} {in} driven active particles \cite{PhysRevE.105.L012101,PhysRevE.110.L052203},
{chemical potentials fueling} kinesin {motion} \cite{liepelt1, liepelt2}, {light-activated transitions in} photo-acids \cite{berton2020thermodynamics}, {thermodynamic forces in chemical reaction networks} \cite{liang2024thermodynamic, busiello2021dissipation, rao2016nonequilibrium}, ATP-driven {pathways in} chaperones and {molecular transporters} \cite{de2014hsp70, flatt2023abc}. Recent studies highlight a distinct phase transition that separates a functional heat engine regime from a ``dud" regime, where the system is purely dissipative \cite{gatien, gatien2, filho2023powerful, mamede2023}. Despite its relevance in biophysics and related fields, the {interplay between biased forces and phase transition regimes} remains poorly understood and {explored only in} a few {examples}.

In this Letter, we demonstrate that collective effects under driving forces yield novel phenomena with {no} analog in traditional equilibrium or nonequilibrium phase transitions, {as evidenced by} three distinct findings. First, we find that order-disorder phase transitions occur at separate points depending on which ordered phase initially dominates, {deviating} from typical mean-field phase {transitions. This} behavior also {differ from} discontinuous nonequilibrium phase transitions known as “generic two-phase coexistence”, where a region, rather than a single point, defines the transition \cite{evans, evans2}. Second, each order-disorder phase transition can fall into a different classification. {For a certain range of parameters,} the ``up''-spin-dominated phase transition is critical {and continous}, while the ``down''-spin-dominated transition is discontinuous. Third, unlike {customary} “up-down” ($Z_2$) symmetry-breaking phenomena \cite{tome2006,noa2019,aguilera2021unifying}, the continuous phase transition {we observe} exhibits critical exponents that differ from the classical result of $\beta = 1/2$. Our findings reveal novel {insights into the properties} of the order parameter {and uncover unique behaviors of key} thermodynamic quantities, {such as} power {and energy} dissipation. We further show that the less dissipative ordered phase -- characterized by lower power fluctuations -- exhibits heat-engine behavior capable of achieving near-optimal efficiency and maximum power. In contrast to some recent studies \cite{pancotti2020speed, erdman2022driving, erdman2023pareto}, our results suggest that these performance improvements arise directly from collective dynamics, without {the need for} complex strategies.\\

{\it Minimal collective model and thermodynamics}---
To illustrate these findings, we propose a minimal model consisting of \( N \) interacting units, where each unit \( j \) is represented by a spin variable \( s_j \) with values \( s_j \in \{-1, 0, 1\} \). Thus, a given microscopic state is defined by the configuration of individual spins, \( s \equiv (s_1, \ldots, s_j, \ldots, s_N) \). The energy of the system takes the simple Ising-like form
\begin{equation} 
E(s) = \frac{\epsilon}{2k} \sum_{(i,j)} s_i s_j + \Delta \sum_{i=1}^N s_i^2,
\label{eqq}
\end{equation}
where \( \epsilon < 0\) {quantifies} the interaction energy between two nearest neighbor units, and each {unit contributes} an individual energy \(0\) or \( \Delta \). The equilibrium version of this model exhibits {a variety of} collective behaviors, including continuous and discontinuous transitions, as well as tricritical points \cite{yeomans1992statistical}.

The system is coupled simultaneously to two thermal baths inducing \emph{opposite} nonconservative (biased) driving forces \cite{gatien, herpich, filho2023powerful, liepelt1, liepelt2, berton2020thermodynamics}. The time evolution of the probability \( p_s(t) \) to observe the system in {the} microstate \(s\) at time \(t\) is governed by the master equation
\[
\dot{p}_s(t) = \sum_{\nu=1}^2 \sum_{s' \neq s} J_{s's}^{(\nu)}(t),
\]
where \(\nu\in\{1,2\}\) labels the thermal bath, \( J_{s's}^{(\nu)}(t) = \omega^{(\nu)}_{s s'} p_{s'}(t) - \omega^{(\nu)}_{s's} p_s(t) \) is the probability flux between microstates \( s \) and \( s' \) due to {the} contact with the \( \nu \)-th bath, and \( \omega^{(\nu)}_{s' s} \) is the corresponding transition rate from \( s \) to \( s' \equiv \{s_1, \ldots, s_{j-1}, \tilde{s}_j, s_{j+1}, \ldots, s_N\} \) (where \( \tilde{s}_j \neq s_j \)). Explicitly, 
\[
\omega^{(\nu)}_{s's} = \Gamma \exp\left(-\frac{\beta_\nu}{2} \left[ E(s') - E(s) + F d_{s's}^{(P, \nu)} \right]\right).
\]
Here, \( F d_{s's}^{(P, \nu)} \) accounts for the driving force, with \( d_{s's}^{(P, \nu)} \in \{-1,+1\} \) depending on the bath and on whether it subjects the spin transition \( s_j \rightarrow \tilde{s}_j \) to a clockwise or a counterclockwise bias. 
Clockwise transitions are defined as those for which an individual state changes following \( -1 \rightarrow 0 \rightarrow +1 \rightarrow -1 \), while counterclockwise transitions correspond to \( +1 \rightarrow 0 \rightarrow -1 \rightarrow +1 \). 
We assume that the cold thermal bath (\(\nu=1\)) favors clockwise transitions, for which we set   
\( d_{s's}^{(P, 1)} = -1 \), {while}
\( d_{s's}^{(P, 1)} = +1 \) for counterclockwise transitions. On the other hand, we assume that the hot thermal bath (\(\nu=2\)) favors counterclockwise transitions, for which \( d_{s's}^{(P, 2)} = -1 \), while for clockwise transitions \( d_{s's}^{(P, 2)} = +1 \). 

The mean power \( \langle \mathcal{P} \rangle \) and mean heat \( \langle \dot{Q}_\nu \rangle \) exchanged with the \( \nu \)-th thermal bath are given by
\[
\langle \dot{Q}_\nu \rangle = \sum_{(s, s')} d_{s's}^{(\dot{Q}_\nu, \nu)} J_{s's}^{(\nu)} \quad \text{and} \quad \langle \mathcal{P} \rangle = -F \sum_{\nu} \sum_{(s, s')} d_{s's}^{(P, \nu)} J_{s's}^{(\nu)},
\]
where \( d_{s's}^{(\dot{Q}_\nu, \nu)} = E(s') - E(s) + d_{s's}^{(P, \nu)} F \) and \( d_{s's}^{(\dot{Q}_\nu, \nu)} = -d_{ss'}^{(\dot{Q}_\nu, \nu)} \). The nonequilibrium steady-state {(NESS),} defined by a time-independent probability set \( \{p^{\text{st}}_s\} \), {has to satisfy} the first {and second law of thermodynamics, respectively,} \( \langle \mathcal{P} \rangle + \langle \dot{Q}_1 \rangle + \langle \dot{Q}_2 \rangle = 0 \) and \( \langle \sigma \rangle = -\beta_1 \langle \dot{Q}_1 \rangle - \beta_2 \langle \dot{Q}_2 \rangle \geq 0 \), {where} \(\langle\sigma\rangle\) is the dissipation rate (or rate of entropy production).

In the nonequilibrium model, interactions give rise to collective effects, where phase transitions emerge as parameters (e.g., \( \epsilon \) or \( \Delta \)) are varied. Similar to the equilibrium model, two phases exist, denoted \( A \) and \( B \), respectively characterized by a predominance of spins \( -1 \) and \( +1 \). Letting \( m = -\langle s_i \rangle \) denote the order parameter, we have \( m > 0 \) for phase \( A \) and \( m < 0 \) for phase \( B \), while \( m = 0 \) indicates independent operation of units. The ``quadrupole moment'' \( q = \langle s_i^2 \rangle \) distinguishes between phases where units predominantly occupy states \( \pm 1 \) (\( q > 0 \)) and where they are in state \( 0 \) (\( q = 0 \)).\\

{\it Emergence of phase transitions with no equilibrium or nonequilibrium analogs}---   
The interplay between biased driving forces and different ordered phases results in 
unique features of phase transitions with no equilibrium analog. {Each ordered phase is associated with a} 
distinct transition {point}, \( X_{1c} \) or \( X_{2c} \) [\( X = \left( \epsilon, \Delta \right) \)] for \( m > 0 \) and \( m < 0 \), respectively. This starkly contrasts to typical phase transitions characterized by spontaneous symmetry breaking. To illustrate these features, we analyze interactions on both square lattice arrangements (coordination number \( k = 4 \)) and the all-to-all version (\( k = N \rightarrow \infty \)), the latter case yielding exact results which we now describe. 

\begin{figure*}[t]
    \centering
    \includegraphics[width=\textwidth]{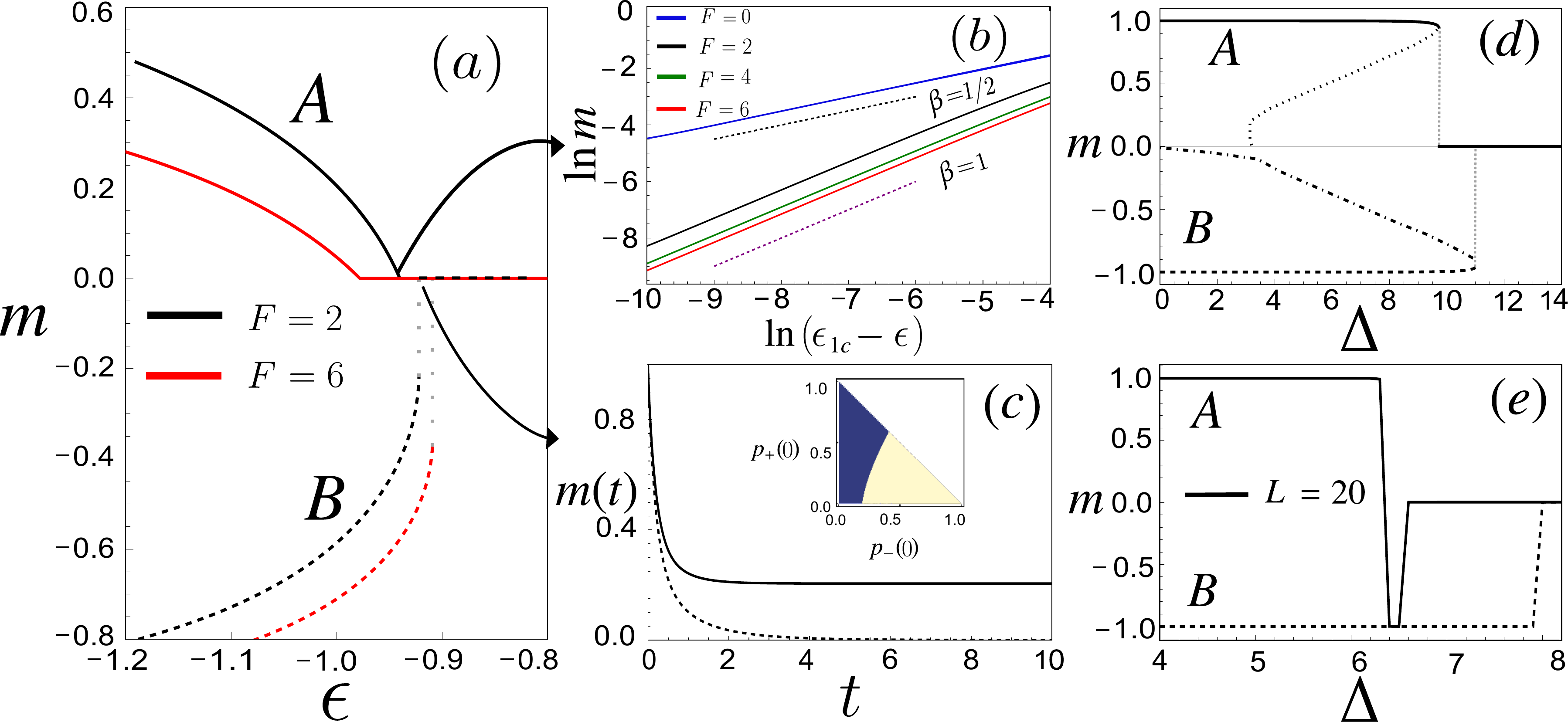}
     \caption{ 
     {(a)} For all-to-all interactions, {we} show the order parameter for different arrangements, with \( m > 0 \) (solid line) and \( m < 0 \) (dashed line), plotted against \( \epsilon \) for various driving strengths. Dotted lines indicate discontinuous phase transitions. {(b)} Critical and discontinuous behaviors for the ordered phases \( A \) and \( B \), respectively; in the discontinuous case, the system evolves to one of two steady-state solutions (solid vs. dotted lines), depending on initial conditions {(see inset in panel (c))}. The right panels demonstrate the robustness of these findings at \( \epsilon = -13 \) (by varying \( \Delta \)) for all-to-all interactions as well as the square lattice (obtained from numerical simulations) in panels {(d) and (e)}, respectively. {In {(d)}, dotted and dotted-dashed lines denote the unstable solutions for the phases A and B respectively. {The order parameter will} converge to the ordered phase [$m>0$ (A) and $m<0$ (B)] {when initial conditions} $m(0)$ lies 
     above (A) and below (B), otherwise it will vanish.} {The sudden jump in panel (e) 
     from $m\approx 1$ to $-1$ is due to finite-size effects.} Unless {stated} otherwise, parameters are \(F=2\),  \(\beta_1 = 2\) and \(\beta_2 = 1\). 
     }
    \label{fig1}
\end{figure*}
Following previous works \cite{gatien,filho2023powerful,herpich,herpich2,mamede2023}, we describe the all-to-all dynamics via the density of units in {one of the} possible individual states, \( n_{i} = \langle N_{i}/N \rangle \), {with} \( i \in \{-,0,+\} \). {Hence,} \(N_\pm\) {represents} the number of units in the local states \(s=\pm 1\) and \(N_0\) {the number of units for which} \(s=0\). 
The dynamics of \(\{n_i\}\) is governed by the master equation 
\[ \dot{n}_i(t) = \sum_{\nu=1}^2 \sum_{j \neq i} J_{ij}^{(\nu)}(t), \]  
where \( J_{ij}^{(\nu)}(t) = \omega^{(\nu)}_{ij} n_{j}(t) - \omega^{(\nu)}_{ji} n_i(t) \) with density-dependent transition rates \(\omega^{(\nu)}_{ji}\) {whose explicit expression is} presented in the Supplemental Material. The order parameter \( m \) and quadrupole moment \( q \) are given by \( m = n_- - n_+ \) and \( q = n_- + n_+ \). In the absence of individual energies (\( \Delta = 0 \)), the steady state of the system corresponds to an ordered phase for \( |\beta_\nu \epsilon| \gg 1 \), with both \( |m| \) and \( q \) close to \(1\). In the disordered phase, \( |\epsilon| < \max\left(|\epsilon_{1c}|,|\epsilon_{2c}|\right)\) at fixed \(\beta_1\) and \(\beta_2\), {and} the steady-state densities are equal, \( n^\mathrm{st}_+ = n^\mathrm{st}_- = n^\mathrm{st}_0 = 1/3 \), yielding \( m = 0\) and \(q = 2/3 \). Notice that, by combining the master equations for \(n_-\) and \(n_+\), it is possible to {derive} master equations for \(m\) and \(q\). In the steady state, \(dm/dt=dq/dt=0\), from which an implicit relation \(q=q(m)\) can be {obtained}.

If there is a continuous phase transition to the disordered state, the steady state values \(\left(m,q\right)\) continuously approach \(\left(0,2/3\right)\) as \(\epsilon<0\) is increased from the ordered phase at fixed \(\beta_1\) and \(\beta_2\). {Therefore, it is possible to} {expand the master equation} in \(m\) {and, taking} into account the implicit relation \(q(m)\), the time evolution of \( m \) {reads as follows}
\begin{equation}
\frac{dm}{dt} \approx a(\epsilon - \epsilon_{1c})m + bm^2 + cm^3 + \ldots,
\label{eq1m}
\end{equation}
where \( \epsilon_{1c} \) is given by  
\[
\epsilon_{1c} = -\frac{e^{\frac{1}{2} F (\beta_1 - \beta_2)} + e^{\frac{1}{2} F (\beta_2 - \beta_1)} + e^{\frac{1}{2} F (\beta_1 + \beta_2)} + e^{\beta_1 F} + e^{\beta_2 F} + 1}{\left(e^{\frac{\beta_1 F}{2}} + e^{\frac{\beta_2 F}{2}}\right) \left(\beta_1 \cosh \left(\frac{\beta_1 F}{2}\right) + \beta_2 \cosh \left(\frac{\beta_2 F}{2}\right)\right)},
\]
while
\[ b = \sinh \left(\frac{F}{4} (\beta_1 - \beta_2)\right) f(\beta_1, \beta_2, F), \]  
with explicit expressions for \( f(\beta_1, \beta_2, F) \) and the coefficients \(a>0\) and \( c \) provided in the Supplemental Material. 

A few observations about Eq.~(\ref{eq1m}) are in order. First, the presence of a term proportional to \( m^2 \) for \( F \neq 0 \) and \( \beta_1 \neq \beta_2 \) leads to the critical behavior \( m \sim a(\epsilon_{1c} - \epsilon)/b \), with a critical exponent \( \beta = 1 \) (not to be confused with the inverse bath temperatures \(\beta_1\) and \(\beta_2\)), markedly different from the standard mean-field behavior in order-disorder phase transitions \( |m| \sim \sqrt{a(\epsilon_{1c} - \epsilon)/c} \), where \( \beta = 1/2 \), obtained when \( F = 0 \) or \( \beta_1 = \beta_2 \). The appearance of a critical exponent \( \beta = 1 \),
illustrated in {Fig.~\ref{fig1}b}, is a consequence of breaking the \( Z_2 \)-``up-down" symmetry which is often present in equilibrium and nonequilibrium phase transitions for mean-field systems \cite{noa2019,tome2006,aguilera2023nonequilibrium}. Two additional distinctions from standard order-disorder phase transitions are that \( \epsilon_{2c} \neq \epsilon_{1c} \) and, for \(\Delta=0\), the classification of the corresponding phase transitions differ. While the transition from  \( m > 0 \) to the disordered phase is continuous, the one from \( m < 0 \) is discontinuous, as shown in Fig.~\ref{fig1}a. {Unlike a critical transition, the latter case features a spinodal region where the system may reach two distinct steady states depending on the initial configuration (see Fig.~\ref{fig1}c).}

The existence of two distinct transition points is robust, as shown in {Figs.~\ref{fig1}d-e, where} we fix the values of \(\beta_1\), \(\beta_2\) and \(\epsilon\) while varying the individual-energy parameter \(\Delta\), both in the all-to-all case {(Fig.~\ref{fig1}d)} and in the square lattice {(Fig.~\ref{fig1}e)}. Notice that now both transitions are discontinuous, {with the jumps in the order parameter and the spinodal regions being more pronounced (dotted and dotted-dashed lines) than for $\Delta=0$}. Notably, {hints of a similar scenario} {have been} observed in a simpler two-state model \cite{gatien,filho2023powerful}, where the \( m > 0 \) phase exhibits a discontinuous phase transition, absent for \( m < 0 \).

\begin{figure*}[t]
    \centering
    \includegraphics[width=\textwidth]{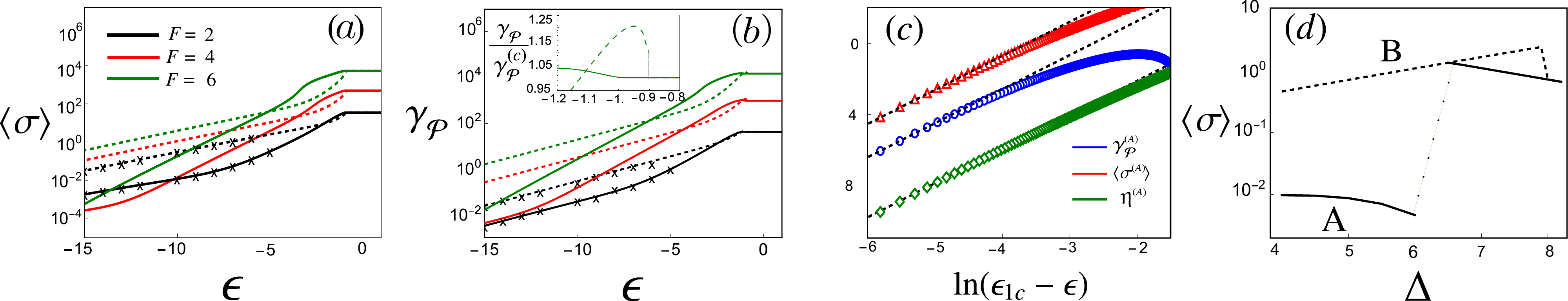}
    \caption{
{(a) Average entropy production as functions of \(\epsilon\) in phases \(A\) (solid lines) and \(B\) (dashed lines). (b) Same as panel (a) but for power fluctuations.} The inset provides a close-up of \(\gamma_{\cal P}\) near the phase transitions {(dashed line indicates phase \(B\))}. These findings are robust beyond the all-to-all interaction case, holding also for square-lattice interactions (data marked by \(\times\)). (c) Log-log plots of the above quantities near the critical point \(\epsilon_{1c}\), with slopes consistent with \(\delta = 2\). {(d)}, entropy production is shown as a function of \(\Delta\) for the same parameters as in Fig.~\ref{fig1}. 
}
    \label{fluc}\end{figure*}

These findings {are also reflected in peculiar thermodynamic properties of this class of systems.} 
In the following, we analyze {the power $\mathcal{P}$ and its fluctuations} \( \gamma_{\mathcal{P}} \equiv \langle \mathcal{P}^2\rangle - \langle\mathcal{P}\rangle^2 \), the efficiency \( \eta = -\langle \mathcal{P} \rangle / \langle \dot{Q}_2 \rangle \), {and the energy} dissipation \( \langle \sigma \rangle \). We use the superscripts \((A)\) and \((B)\) to distinguish between quantities associated with phases \(A\) (\(m>0\)) and \(B\) (\(m<0\)). 
{Following the same procedure as above,} by expanding \( \langle \sigma^{(A)} \rangle \) and \( \gamma_{\mathcal{P}}^{(A)} \) in terms of the order parameter near \( \epsilon_{1c} \), we arrive at {the following expressions}:
\begin{eqnarray*}
    \langle \sigma^{(A)} \rangle &\sim& \langle \sigma_c \rangle + c_{\sigma} m^2 + \cdots \\
    \gamma_{\mathcal{P}}^{(A)} &\sim& \gamma_{\mathcal{P}}^{(c)} + c_v m^2 + \cdots, 
\end{eqnarray*}
where \( \langle \sigma_c \rangle = 2F \left[ \beta_1 \sinh(\beta_1 F / 2) + \beta_2 \sinh(\beta_2 F / 2) \right] \) and \( \gamma_{\mathcal{P}}^{(c)} = 2F^2 \left[ \cosh(\beta_1 F / 2) + \cosh(\beta_2 F / 2) \right] \) respectively denote entropy production and power variance for \( \epsilon_1 \ge \epsilon_{1c} \). The coefficients \( c_{\sigma} \) and \( c_v \) are provided in the Supplemental Material. {This dependence on the} order parameter {implies another scaling behavior, i.e.,} \( \langle \sigma^{(A)} \rangle - \langle \sigma_c \rangle \sim a c_{\sigma} (\epsilon_{1c} - \epsilon)^{\delta} / b \) and \( \gamma_{\mathcal{P}}^{(A)} - \gamma_{\mathcal{P}}^{2(c)} \sim a c_v (\epsilon_{1c} - \epsilon)^{\delta} / b \), with \( \delta = 2\beta = 2 \). {Average entropy production and power fluctuations as a function of $\epsilon$ are in Fig.~\ref{fluc}a-b.}

The system’s performance, particularly its efficiency \(\eta\), also has {a scaling behavior characterizing phase (A).} 
To examine this, we expand \(\langle \dot Q^{(A)}_i\rangle\) around \(\epsilon_{1c}\), yielding \(\langle \dot Q^{(A)}_i\rangle \sim \langle \dot Q_i^{(c)}\rangle + c_{q_i} m^2 + \cdots\), where \(\langle \dot Q_i^{(c)}\rangle = -2F \sinh(\beta_i F / 2)\) and the coefficients \(c_{q_1}\) and \(c_{q_2}\) are listed in the Supplemental Material. Using the expressions for \(\langle \dot Q^{(A)}_2\rangle\) and \(\langle {\cal P}^{(A)}\rangle\), we derive an asymptotic form of {the efficiency} \(\eta^{(A)}\),
\[
\eta^{(A)} \sim \eta^{(c)} + \frac{1}{\langle \dot{Q}_2^{(c)} \rangle} \left(c_{q_1} - \frac{c_{q_2} \langle \dot{Q}_1^{(c)} \rangle}{\langle \dot{Q}_2^{(c)} \rangle}\right)(\epsilon_{1c} - \epsilon)^{\delta},
\]
where \(\eta^{(c)} = 1 + \sinh(\beta_1 F / 2) / \sinh(\beta_2 F / 2)\) and \(\delta = 2\beta\). Thus, the efficiency exhibits the same critical exponent as entropy production, power, {and power} fluctuations, as illustrated in {Fig.~\ref{fluc}c}. Although closed-form expressions are unavailable near \(\epsilon_{2c}\), Fig.~\ref{fluc}b ({dashed line in the inset) shows} that the discontinuity in the order parameter {can be equally found in power fluctuations \(\gamma_{\cal P}\).} Moreover, when analyzing the system into the ordered phases, we notice that phase \(A\) (\(m > 0\)) exhibits notably {lower dissipation and power fluctuations than phase \(B\) (\(m < 0\)), as shown in Figs.~\ref{fluc}a-b.}

To {gain more intuition on the system's behavior}, we develop a phenomenological model valid when \(|m| \approx 1\), {using an approach analogous to that introduced in \cite{filho2023powerful}}. Under this condition, the state densities are approximately given by \(n_- = 1\) (\(n_- = 0\)) and \(n_+ = 0\) (\(n_+ = 1\)) for phase \(A\) (\(B\)). Applying the spanning-tree method \cite{schn}, we derive approximate steady-state solutions, depending only on model parameters (details in the Supplemental Material). For \(|\beta_\nu(\epsilon + \Delta)| \gg 1\), the effective expressions for the entropy production are given by
\[
\langle \sigma_{\text{eff}}^{(A)} \rangle \approx -e^{\frac{1}{2} \beta_2 (\Delta - F + \epsilon)} [F (\beta_1 + \beta_2) + (\beta_1 - \beta_2)(\Delta + \epsilon)]
\]
and
\[
\langle \sigma_{\text{eff}}^{(B)} \rangle \approx e^{\frac{\beta_2}{2} (\Delta + F + \epsilon)} [F (\beta_1 + \beta_2) - (\beta_1 - \beta_2)(\Delta + \epsilon)]
\]
for phases \(A\) and \(B\), respectively. Notice that \(\langle \sigma_{\text{eff}}^{(A)}\rangle < \langle \sigma_{\text{eff}}^{(B)}\rangle\). Similarly, expressions for the power {fluctuations} \(\gamma_{\cal P}\) are obtained via the large-deviation method \cite{touchette2009large,kumar2011thermodynamics}, with leading terms given by \(\gamma^{(A)}_{\cal P} \approx 4F^2 \left[e^{\frac{\beta_1}{2} (\epsilon + F)} + e^{\frac{\beta_2}{2} (\epsilon - F)}\right]\) and \(\gamma^{(B)}_{\cal P} \approx 4F^2 e^{\frac{\beta_2}{2} (\epsilon + F)}\) for phases \(A\) and \(B\), respectively. Here again, \(\gamma^{(A)}_{\cal P} < \gamma^{(B)}_{\cal P}\), consistent with phase \(A\) exhibiting lower power fluctuations than phase \(B\). {These results extend beyond the all-to-all scenario} to square-lattice topologies (see data marked by the symbols \(\times\) in panels a-b of Fig.~\ref{fluc}) 

Also, in Fig.~\ref{fluc}d, {we show the behavior of} \(\langle \sigma \rangle\) {in} the square-lattice as \(\Delta\) varies. {Again,} \(\langle \sigma^{(A)} \rangle\) (solid lines) {remains} lower than \(\langle \sigma^{(B)} \rangle\) (dashed lines). 
The entropy production, like the order parameter, captures {again the} phase transitions occurring at distinct \(\Delta_{1c} \neq \Delta_{2c}\).

\begin{figure*}[t]
    \centering
    \includegraphics[width=\textwidth]{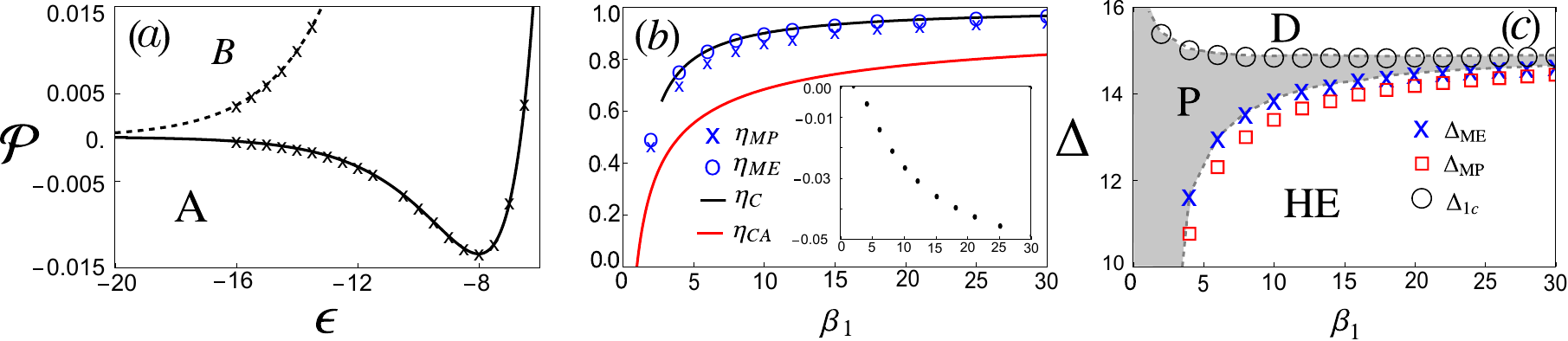}
    \caption{{(a)} Power ${\cal P}$ as functions of \(\epsilon\) in phases \(A\) (solid lines) and \(B\) (dashed lines) for the same parameters {as} Fig.~\ref{fig1}. Symbols {indicate} the square-lattice {case}.
    {(b)} $\eta_{MP}$ (symbols $\times$) and $\eta_{ME}$ (symbols $\circ$) versus $\beta_1$ for $\epsilon = -20$ and $F=5$. 
    For comparison, the continuous {black and red} represent the corresponding Carnot $\eta_c$ and the Curzon-Ahlborn efficiency $\eta_{CA}$, respectively. 
    The inset shows the corresponding maximum power values, ${\cal P}_{MP}$. {(c) The convergence of $\Delta_{MP}$ and $\Delta_{ME}$ to $\Delta_{1c}$ is shown, as a function of $\beta_1$ for the same parameters as in panel (b). Dashed curves here} separate the heat-engine (HE) and dud (D) regimes from the pump {behavior (grey area)}. In all cases, $\beta_2 = 1$.}
    \label{fig3}
\end{figure*}
    
{\it Phase transitions and heat engines operating on the verge of maximum power and maximum efficiency}---
Systems operating collectively can offer significant advantages, such as reducing dissipated work \cite{rolandi,Meibohm24} and optimizing efficiency {and/or} power through tailored {internal structures} \cite{gatien,filho2023powerful,liang2023minimal}. {Here, we notice a crucial difference in system behavior emerging from the existence of the two phases. Indeed, heat engine operations only appear} for \( m > 0 \) (phase \(A\)) and not for \( m < 0 \) (phase \(B\)), as shown in Fig.~\ref{fig3}a. {The} phenomenological description {employed above} sheds {more} light on {this behavior, by investigating the expressions for} \( \langle \mathcal{P} \rangle \) and \( \eta \). When \( |\beta_\nu(\epsilon + \Delta)| \gg 1 \), {these quantities read}
\begin{equation}
\langle \mathcal{P}_{\rm eff}^{(A)}\rangle = \frac{2 F \left(e^{\frac{1}{2} \beta_1 (\Delta + \epsilon + F)} - e^{\frac{1}{2} \beta_2 (\Delta + \epsilon - F)}\right)}{e^{\frac{1}{2} ((\beta_1 + \beta_2) (\Delta + \epsilon) + F (\beta_1 - \beta_2))} + 1},
\label{prodef}
\end{equation}
and
\begin{equation}
\eta^{(A)}_{\rm eff} = \frac{2 F}{F - \epsilon - \Delta} \left(e^{\frac{1}{2} \{(\beta_1 + \beta_2) (\Delta + \epsilon) + F (\beta_1 - \beta_2)\}} + 1\right),
\label{poweref}
\end{equation}
for phase \(A\). Similar expressions for phase \(B\) show that \( \langle \mathcal{P}_{\rm eff}^{(B)} \rangle > 0 \) and \( \eta^{(B)}_{\rm eff} < 0 \). Phase \(A\) uniquely allows a region where the heat engine operates close to maximum power and efficiency, {as shown in Fig.~\ref{fig3}b and Supplemental Material.} While the second law of thermodynamics prevents ideal efficiency at finite power, the interactions and individual energy distributions allow {for the} maximization of power and efficiency at values of \(\Delta\) denoted by \(\Delta_{MP}\) and \(\Delta_{ME}\), respectively, with {the} other parameters {held} fixed. These values are close to each other, enabling near-ideal performance. They converge toward the ideal efficiency \( \eta_c = 1 - \beta_2 / \beta_1 \) as \( \beta_1 \) increases. These findings hold not only for all-to-all interactions but also in square lattice configurations (marked by \(\times\) symbols), without requiring specialized optimization procedures, {such as reward functions that balance power and efficiency. }

Finally, we address the link between optimization and phase transitions. Although \(\Delta_{MP}\) and \(\Delta_{ME}\) don’t necessarily coincide, both converge to the discontinuous transition value \( \Delta = \Delta_{1c} \) as temperature differences increase {(see Fig.~\ref{fig3}c)}, revealing a connection between ideal operational conditions and phase transitions.

{\it Conclusions}--- Our results reveal a class of collective behaviors under driven forces, with a unique phase transition characterized by two distinct transition points. These findings highlight the emergence of ordered phases with contrasting behaviors, where phase \(A\), the less dissipative phase, exhibits lower power fluctuations and supports nearly ideal performance under optimal energy settings. Our work opens avenues for {a} deeper understanding of {sheer} nonequilibrium {phenomena driving collective operation and instantiating efficient heat engines at the nanoscale}. Future investigations could explore more complex systems, such as those with energetic frustration, which frequently appear in nature and could yield further insights {into this field}.

{\it Acknowledgments}---We acknowledge the financial support
from brazilian agencies CNPq and FAPESP under grants No. 2021/03372-2, No. 2022/16192-5,
No. 2022/15453-0, and No. 2024/03763-0.

\bibliography{refs}

\begin{thebibliography}{42}%
\makeatletter
\providecommand \@ifxundefined [1]{%
 \@ifx{#1\undefined}
}%
\providecommand \@ifnum [1]{%
 \ifnum #1\expandafter \@firstoftwo
 \else \expandafter \@secondoftwo
 \fi
}%
\providecommand \@ifx [1]{%
 \ifx #1\expandafter \@firstoftwo
 \else \expandafter \@secondoftwo
 \fi
}%
\providecommand \natexlab [1]{#1}%
\providecommand \enquote  [1]{``#1''}%
\providecommand \bibnamefont  [1]{#1}%
\providecommand \bibfnamefont [1]{#1}%
\providecommand \citenamefont [1]{#1}%
\providecommand \href@noop [0]{\@secondoftwo}%
\providecommand \href [0]{\begingroup \@sanitize@url \@href}%
\providecommand \@href[1]{\@@startlink{#1}\@@href}%
\providecommand \@@href[1]{\endgroup#1\@@endlink}%
\providecommand \@sanitize@url [0]{\catcode `\\12\catcode `\$12\catcode
  `\&12\catcode `\#12\catcode `\^12\catcode `\_12\catcode `\%12\relax}%
\providecommand \@@startlink[1]{}%
\providecommand \@@endlink[0]{}%
\providecommand \url  [0]{\begingroup\@sanitize@url \@url }%
\providecommand \@url [1]{\endgroup\@href {#1}{\urlprefix }}%
\providecommand \urlprefix  [0]{URL }%
\providecommand \Eprint [0]{\href }%
\providecommand \doibase [0]{http://dx.doi.org/}%
\providecommand \selectlanguage [0]{\@gobble}%
\providecommand \bibinfo  [0]{\@secondoftwo}%
\providecommand \bibfield  [0]{\@secondoftwo}%
\providecommand \translation [1]{[#1]}%
\providecommand \BibitemOpen [0]{}%
\providecommand \bibitemStop [0]{}%
\providecommand \bibitemNoStop [0]{.\EOS\space}%
\providecommand \EOS [0]{\spacefactor3000\relax}%
\providecommand \BibitemShut  [1]{\csname bibitem#1\endcsname}%
\let\auto@bib@innerbib\@empty
\bibitem [{\citenamefont {Tom\'e}\ and\ \citenamefont
  {de~Oliveira}(2012)}]{tome2012}%
  \BibitemOpen
  \bibfield  {author} {\bibinfo {author} {\bibfnamefont {T.}~\bibnamefont
  {Tom\'e}}\ and\ \bibinfo {author} {\bibfnamefont {M.~J.}\ \bibnamefont
  {de~Oliveira}},\ }\href {\doibase 10.1103/PhysRevLett.108.020601} {\bibfield
  {journal} {\bibinfo  {journal} {Phys. Rev. Lett.}\ }\textbf {\bibinfo
  {volume} {108}},\ \bibinfo {pages} {020601} (\bibinfo {year}
  {2012})}\BibitemShut {NoStop}%
\bibitem [{\citenamefont {Noa}\ \emph {et~al.}(2019)\citenamefont {Noa},
  \citenamefont {Harunari}, \citenamefont {de~Oliveira},\ and\ \citenamefont
  {Fiore}}]{noa2019}%
  \BibitemOpen
  \bibfield  {author} {\bibinfo {author} {\bibfnamefont {C.~E.~F.}\
  \bibnamefont {Noa}}, \bibinfo {author} {\bibfnamefont {P.~E.}\ \bibnamefont
  {Harunari}}, \bibinfo {author} {\bibfnamefont {M.~J.}\ \bibnamefont
  {de~Oliveira}}, \ and\ \bibinfo {author} {\bibfnamefont {C.~E.}\ \bibnamefont
  {Fiore}},\ }\href {\doibase 10.1103/PhysRevE.100.012104} {\bibfield
  {journal} {\bibinfo  {journal} {Phys. Rev. E}\ }\textbf {\bibinfo {volume}
  {100}},\ \bibinfo {pages} {012104} (\bibinfo {year} {2019})}\BibitemShut
  {NoStop}%
\bibitem [{\citenamefont {Aguilera}\ \emph {et~al.}(2023)\citenamefont
  {Aguilera}, \citenamefont {Igarashi},\ and\ \citenamefont
  {Shimazaki}}]{aguilera2023nonequilibrium}%
  \BibitemOpen
  \bibfield  {author} {\bibinfo {author} {\bibfnamefont {M.}~\bibnamefont
  {Aguilera}}, \bibinfo {author} {\bibfnamefont {M.}~\bibnamefont {Igarashi}},
  \ and\ \bibinfo {author} {\bibfnamefont {H.}~\bibnamefont {Shimazaki}},\
  }\href@noop {} {\bibfield  {journal} {\bibinfo  {journal} {Nature
  Communications}\ }\textbf {\bibinfo {volume} {14}},\ \bibinfo {pages} {3685}
  (\bibinfo {year} {2023})}\BibitemShut {NoStop}%
\bibitem [{\citenamefont {Loos}\ \emph {et~al.}(2023)\citenamefont {Loos},
  \citenamefont {Klapp},\ and\ \citenamefont {Martynec}}]{loos2023long}%
  \BibitemOpen
  \bibfield  {author} {\bibinfo {author} {\bibfnamefont {S.~A.}\ \bibnamefont
  {Loos}}, \bibinfo {author} {\bibfnamefont {S.~H.}\ \bibnamefont {Klapp}}, \
  and\ \bibinfo {author} {\bibfnamefont {T.}~\bibnamefont {Martynec}},\
  }\href@noop {} {\bibfield  {journal} {\bibinfo  {journal} {Physical review
  letters}\ }\textbf {\bibinfo {volume} {130}},\ \bibinfo {pages} {198301}
  (\bibinfo {year} {2023})}\BibitemShut {NoStop}%
\bibitem [{\citenamefont {Herpich}\ \emph {et~al.}(2018)\citenamefont
  {Herpich}, \citenamefont {Thingna},\ and\ \citenamefont
  {Esposito}}]{herpich}%
  \BibitemOpen
  \bibfield  {author} {\bibinfo {author} {\bibfnamefont {T.}~\bibnamefont
  {Herpich}}, \bibinfo {author} {\bibfnamefont {J.}~\bibnamefont {Thingna}}, \
  and\ \bibinfo {author} {\bibfnamefont {M.}~\bibnamefont {Esposito}},\ }\href
  {\doibase 10.1103/PhysRevX.8.031056} {\bibfield  {journal} {\bibinfo
  {journal} {Phys. Rev. X}\ }\textbf {\bibinfo {volume} {8}},\ \bibinfo {pages}
  {031056} (\bibinfo {year} {2018})}\BibitemShut {NoStop}%
\bibitem [{\citenamefont {Herpich}\ and\ \citenamefont
  {Esposito}(2019)}]{herpich2}%
  \BibitemOpen
  \bibfield  {author} {\bibinfo {author} {\bibfnamefont {T.}~\bibnamefont
  {Herpich}}\ and\ \bibinfo {author} {\bibfnamefont {M.}~\bibnamefont
  {Esposito}},\ }\href {\doibase 10.1103/PhysRevE.99.022135} {\bibfield
  {journal} {\bibinfo  {journal} {Phys. Rev. E}\ }\textbf {\bibinfo {volume}
  {99}},\ \bibinfo {pages} {022135} (\bibinfo {year} {2019})}\BibitemShut
  {NoStop}%
\bibitem [{\citenamefont {Filho}\ \emph {et~al.}(2023)\citenamefont {Filho},
  \citenamefont {For\~ao}, \citenamefont {Busiello}, \citenamefont {Cleuren},\
  and\ \citenamefont {Fiore}}]{filho2023powerful}%
  \BibitemOpen
  \bibfield  {author} {\bibinfo {author} {\bibfnamefont {F.~S.}\ \bibnamefont
  {Filho}}, \bibinfo {author} {\bibfnamefont {G.~A.~L.}\ \bibnamefont
  {For\~ao}}, \bibinfo {author} {\bibfnamefont {D.~M.}\ \bibnamefont
  {Busiello}}, \bibinfo {author} {\bibfnamefont {B.}~\bibnamefont {Cleuren}}, \
  and\ \bibinfo {author} {\bibfnamefont {C.~E.}\ \bibnamefont {Fiore}},\ }\href
  {\doibase 10.1103/PhysRevResearch.5.043067} {\bibfield  {journal} {\bibinfo
  {journal} {Phys. Rev. Res.}\ }\textbf {\bibinfo {volume} {5}},\ \bibinfo
  {pages} {043067} (\bibinfo {year} {2023})}\BibitemShut {NoStop}%
\bibitem [{\citenamefont {Mamede}\ \emph {et~al.}(2023)\citenamefont {Mamede},
  \citenamefont {Proesmans},\ and\ \citenamefont {Fiore}}]{mamede2023}%
  \BibitemOpen
  \bibfield  {author} {\bibinfo {author} {\bibfnamefont {I.~N.}\ \bibnamefont
  {Mamede}}, \bibinfo {author} {\bibfnamefont {K.}~\bibnamefont {Proesmans}}, \
  and\ \bibinfo {author} {\bibfnamefont {C.~E.}\ \bibnamefont {Fiore}},\ }\href
  {\doibase 10.1103/PhysRevResearch.5.043278} {\bibfield  {journal} {\bibinfo
  {journal} {Phys. Rev. Res.}\ }\textbf {\bibinfo {volume} {5}},\ \bibinfo
  {pages} {043278} (\bibinfo {year} {2023})}\BibitemShut {NoStop}%
\bibitem [{\citenamefont {Feyisa}\ and\ \citenamefont {Jen}(2024)}]{Gashu024}%
  \BibitemOpen
  \bibfield  {author} {\bibinfo {author} {\bibfnamefont {C.~G.}\ \bibnamefont
  {Feyisa}}\ and\ \bibinfo {author} {\bibfnamefont {H.~H.}\ \bibnamefont
  {Jen}},\ }\href {\doibase 10.1088/1367-2630/ad7c74} {\bibfield  {journal}
  {\bibinfo  {journal} {New Journal of Physics}\ }\textbf {\bibinfo {volume}
  {26}},\ \bibinfo {pages} {093039} (\bibinfo {year} {2024})}\BibitemShut
  {NoStop}%
\bibitem [{\citenamefont {Bettmann}\ \emph {et~al.}(2023)\citenamefont
  {Bettmann}, \citenamefont {Kewming},\ and\ \citenamefont
  {Goold}}]{PhysRevE.107.044102}%
  \BibitemOpen
  \bibfield  {author} {\bibinfo {author} {\bibfnamefont {L.~P.}\ \bibnamefont
  {Bettmann}}, \bibinfo {author} {\bibfnamefont {M.~J.}\ \bibnamefont
  {Kewming}}, \ and\ \bibinfo {author} {\bibfnamefont {J.}~\bibnamefont
  {Goold}},\ }\href {\doibase 10.1103/PhysRevE.107.044102} {\bibfield
  {journal} {\bibinfo  {journal} {Phys. Rev. E}\ }\textbf {\bibinfo {volume}
  {107}},\ \bibinfo {pages} {044102} (\bibinfo {year} {2023})}\BibitemShut
  {NoStop}%
\bibitem [{\citenamefont {Liang}\ \emph {et~al.}(2023)\citenamefont {Liang},
  \citenamefont {Ma}, \citenamefont {Busiello},\ and\ \citenamefont
  {Rios}}]{liang2023minimal}%
  \BibitemOpen
  \bibfield  {author} {\bibinfo {author} {\bibfnamefont {S.}~\bibnamefont
  {Liang}}, \bibinfo {author} {\bibfnamefont {Y.-H.}\ \bibnamefont {Ma}},
  \bibinfo {author} {\bibfnamefont {D.~M.}\ \bibnamefont {Busiello}}, \ and\
  \bibinfo {author} {\bibfnamefont {P.~D.~L.}\ \bibnamefont {Rios}},\
  }\href@noop {} {\bibfield  {journal} {\bibinfo  {journal} {arXiv preprint
  arXiv:2312.02323}\ } (\bibinfo {year} {2023})}\BibitemShut {NoStop}%
\bibitem [{\citenamefont {Rolandi}\ \emph {et~al.}(2023)\citenamefont
  {Rolandi}, \citenamefont {Abiuso},\ and\ \citenamefont
  {Perarnau-Llobet}}]{rolandi}%
  \BibitemOpen
  \bibfield  {author} {\bibinfo {author} {\bibfnamefont {A.}~\bibnamefont
  {Rolandi}}, \bibinfo {author} {\bibfnamefont {P.}~\bibnamefont {Abiuso}}, \
  and\ \bibinfo {author} {\bibfnamefont {M.}~\bibnamefont {Perarnau-Llobet}},\
  }\href {\doibase 10.1103/PhysRevLett.131.210401} {\bibfield  {journal}
  {\bibinfo  {journal} {Phys. Rev. Lett.}\ }\textbf {\bibinfo {volume} {131}},\
  \bibinfo {pages} {210401} (\bibinfo {year} {2023})}\BibitemShut {NoStop}%
\bibitem [{\citenamefont {Campisi}\ and\ \citenamefont
  {Fazio}(2016)}]{campisi2016power}%
  \BibitemOpen
  \bibfield  {author} {\bibinfo {author} {\bibfnamefont {M.}~\bibnamefont
  {Campisi}}\ and\ \bibinfo {author} {\bibfnamefont {R.}~\bibnamefont
  {Fazio}},\ }\href@noop {} {\bibfield  {journal} {\bibinfo  {journal} {Nature
  communications}\ }\textbf {\bibinfo {volume} {7}},\ \bibinfo {pages} {1}
  (\bibinfo {year} {2016})}\BibitemShut {NoStop}%
\bibitem [{\citenamefont {Souza}\ \emph {et~al.}(2022)\citenamefont {Souza},
  \citenamefont {Manzano}, \citenamefont {Fazio},\ and\ \citenamefont
  {Iemini}}]{souza2022collective}%
  \BibitemOpen
  \bibfield  {author} {\bibinfo {author} {\bibfnamefont {L.~d.~S.}\
  \bibnamefont {Souza}}, \bibinfo {author} {\bibfnamefont {G.}~\bibnamefont
  {Manzano}}, \bibinfo {author} {\bibfnamefont {R.}~\bibnamefont {Fazio}}, \
  and\ \bibinfo {author} {\bibfnamefont {F.}~\bibnamefont {Iemini}},\
  }\href@noop {} {\bibfield  {journal} {\bibinfo  {journal} {Physical Review
  E}\ }\textbf {\bibinfo {volume} {106}},\ \bibinfo {pages} {014143} (\bibinfo
  {year} {2022})}\BibitemShut {NoStop}%
\bibitem [{\citenamefont {Vroylandt}\ \emph {et~al.}(2017)\citenamefont
  {Vroylandt}, \citenamefont {Esposito},\ and\ \citenamefont
  {Verley}}]{gatien}%
  \BibitemOpen
  \bibfield  {author} {\bibinfo {author} {\bibfnamefont {H.}~\bibnamefont
  {Vroylandt}}, \bibinfo {author} {\bibfnamefont {M.}~\bibnamefont {Esposito}},
  \ and\ \bibinfo {author} {\bibfnamefont {G.}~\bibnamefont {Verley}},\ }\href
  {\doibase 10.1209/0295-5075/120/30009} {\bibfield  {journal} {\bibinfo
  {journal} {{EPL} (Europhysics Letters)}\ }\textbf {\bibinfo {volume} {120}},\
  \bibinfo {pages} {30009} (\bibinfo {year} {2017})}\BibitemShut {NoStop}%
\bibitem [{\citenamefont {Vroylandt}\ \emph {et~al.}(2020)\citenamefont
  {Vroylandt}, \citenamefont {Esposito},\ and\ \citenamefont
  {Verley}}]{gatien2}%
  \BibitemOpen
  \bibfield  {author} {\bibinfo {author} {\bibfnamefont {H.}~\bibnamefont
  {Vroylandt}}, \bibinfo {author} {\bibfnamefont {M.}~\bibnamefont {Esposito}},
  \ and\ \bibinfo {author} {\bibfnamefont {G.}~\bibnamefont {Verley}},\ }\href
  {\doibase 10.1103/PhysRevLett.124.250603} {\bibfield  {journal} {\bibinfo
  {journal} {Phys. Rev. Lett.}\ }\textbf {\bibinfo {volume} {124}},\ \bibinfo
  {pages} {250603} (\bibinfo {year} {2020})}\BibitemShut {NoStop}%
\bibitem [{\citenamefont {Valani}(2022)}]{PhysRevE.105.L012101}%
  \BibitemOpen
  \bibfield  {author} {\bibinfo {author} {\bibfnamefont {R.~N.}\ \bibnamefont
  {Valani}},\ }\href {\doibase 10.1103/PhysRevE.105.L012101} {\bibfield
  {journal} {\bibinfo  {journal} {Phys. Rev. E}\ }\textbf {\bibinfo {volume}
  {105}},\ \bibinfo {pages} {L012101} (\bibinfo {year} {2022})}\BibitemShut
  {NoStop}%
\bibitem [{\citenamefont {Valani}\ and\ \citenamefont
  {Dandogbessi}(2024)}]{PhysRevE.110.L052203}%
  \BibitemOpen
  \bibfield  {author} {\bibinfo {author} {\bibfnamefont {R.~N.}\ \bibnamefont
  {Valani}}\ and\ \bibinfo {author} {\bibfnamefont {B.~S.}\ \bibnamefont
  {Dandogbessi}},\ }\href {\doibase 10.1103/PhysRevE.110.L052203} {\bibfield
  {journal} {\bibinfo  {journal} {Phys. Rev. E}\ }\textbf {\bibinfo {volume}
  {110}},\ \bibinfo {pages} {L052203} (\bibinfo {year} {2024})}\BibitemShut
  {NoStop}%
\bibitem [{\citenamefont {Liepelt}\ and\ \citenamefont
  {Lipowsky}(2007)}]{liepelt1}%
  \BibitemOpen
  \bibfield  {author} {\bibinfo {author} {\bibfnamefont {S.}~\bibnamefont
  {Liepelt}}\ and\ \bibinfo {author} {\bibfnamefont {R.}~\bibnamefont
  {Lipowsky}},\ }\href@noop {} {\bibfield  {journal} {\bibinfo  {journal}
  {Phys. Rev. Lett.}\ }\textbf {\bibinfo {volume} {98}},\ \bibinfo {pages}
  {258102} (\bibinfo {year} {2007})}\BibitemShut {NoStop}%
\bibitem [{\citenamefont {Liepelt}\ and\ \citenamefont
  {Lipowsky}(2009)}]{liepelt2}%
  \BibitemOpen
  \bibfield  {author} {\bibinfo {author} {\bibfnamefont {S.}~\bibnamefont
  {Liepelt}}\ and\ \bibinfo {author} {\bibfnamefont {R.}~\bibnamefont
  {Lipowsky}},\ }\href {\doibase 10.1103/PhysRevE.79.011917} {\bibfield
  {journal} {\bibinfo  {journal} {Phys. Rev. E}\ }\textbf {\bibinfo {volume}
  {79}},\ \bibinfo {pages} {011917} (\bibinfo {year} {2009})}\BibitemShut
  {NoStop}%
\bibitem [{\citenamefont {Berton}\ \emph {et~al.}(2020)\citenamefont {Berton},
  \citenamefont {Busiello}, \citenamefont {Zamuner}, \citenamefont {Solari},
  \citenamefont {Scopelliti}, \citenamefont {Fadaei-Tirani}, \citenamefont
  {Severin},\ and\ \citenamefont {Pezzato}}]{berton2020thermodynamics}%
  \BibitemOpen
  \bibfield  {author} {\bibinfo {author} {\bibfnamefont {C.}~\bibnamefont
  {Berton}}, \bibinfo {author} {\bibfnamefont {D.~M.}\ \bibnamefont
  {Busiello}}, \bibinfo {author} {\bibfnamefont {S.}~\bibnamefont {Zamuner}},
  \bibinfo {author} {\bibfnamefont {E.}~\bibnamefont {Solari}}, \bibinfo
  {author} {\bibfnamefont {R.}~\bibnamefont {Scopelliti}}, \bibinfo {author}
  {\bibfnamefont {F.}~\bibnamefont {Fadaei-Tirani}}, \bibinfo {author}
  {\bibfnamefont {K.}~\bibnamefont {Severin}}, \ and\ \bibinfo {author}
  {\bibfnamefont {C.}~\bibnamefont {Pezzato}},\ }\href@noop {} {\bibfield
  {journal} {\bibinfo  {journal} {Chemical Science}\ }\textbf {\bibinfo
  {volume} {11}},\ \bibinfo {pages} {8457} (\bibinfo {year}
  {2020})}\BibitemShut {NoStop}%
\bibitem [{\citenamefont {Liang}\ \emph {et~al.}(2024)\citenamefont {Liang},
  \citenamefont {De~Los~Rios},\ and\ \citenamefont
  {Busiello}}]{liang2024thermodynamic}%
  \BibitemOpen
  \bibfield  {author} {\bibinfo {author} {\bibfnamefont {S.}~\bibnamefont
  {Liang}}, \bibinfo {author} {\bibfnamefont {P.}~\bibnamefont {De~Los~Rios}},
  \ and\ \bibinfo {author} {\bibfnamefont {D.~M.}\ \bibnamefont {Busiello}},\
  }\href@noop {} {\bibfield  {journal} {\bibinfo  {journal} {Physical Review
  Letters}\ }\textbf {\bibinfo {volume} {132}},\ \bibinfo {pages} {228402}
  (\bibinfo {year} {2024})}\BibitemShut {NoStop}%
\bibitem [{\citenamefont {Busiello}\ \emph {et~al.}(2021)\citenamefont
  {Busiello}, \citenamefont {Liang}, \citenamefont {Piazza},\ and\
  \citenamefont {De~Los~Rios}}]{busiello2021dissipation}%
  \BibitemOpen
  \bibfield  {author} {\bibinfo {author} {\bibfnamefont {D.~M.}\ \bibnamefont
  {Busiello}}, \bibinfo {author} {\bibfnamefont {S.}~\bibnamefont {Liang}},
  \bibinfo {author} {\bibfnamefont {F.}~\bibnamefont {Piazza}}, \ and\ \bibinfo
  {author} {\bibfnamefont {P.}~\bibnamefont {De~Los~Rios}},\ }\href@noop {}
  {\bibfield  {journal} {\bibinfo  {journal} {Communications Chemistry}\
  }\textbf {\bibinfo {volume} {4}},\ \bibinfo {pages} {16} (\bibinfo {year}
  {2021})}\BibitemShut {NoStop}%
\bibitem [{\citenamefont {Rao}\ and\ \citenamefont
  {Esposito}(2016)}]{rao2016nonequilibrium}%
  \BibitemOpen
  \bibfield  {author} {\bibinfo {author} {\bibfnamefont {R.}~\bibnamefont
  {Rao}}\ and\ \bibinfo {author} {\bibfnamefont {M.}~\bibnamefont {Esposito}},\
  }\href@noop {} {\bibfield  {journal} {\bibinfo  {journal} {Physical Review
  X}\ }\textbf {\bibinfo {volume} {6}},\ \bibinfo {pages} {041064} (\bibinfo
  {year} {2016})}\BibitemShut {NoStop}%
\bibitem [{\citenamefont {De~Los~Rios}\ and\ \citenamefont
  {Barducci}(2014)}]{de2014hsp70}%
  \BibitemOpen
  \bibfield  {author} {\bibinfo {author} {\bibfnamefont {P.}~\bibnamefont
  {De~Los~Rios}}\ and\ \bibinfo {author} {\bibfnamefont {A.}~\bibnamefont
  {Barducci}},\ }\href@noop {} {\bibfield  {journal} {\bibinfo  {journal}
  {Elife}\ }\textbf {\bibinfo {volume} {3}},\ \bibinfo {pages} {e02218}
  (\bibinfo {year} {2014})}\BibitemShut {NoStop}%
\bibitem [{\citenamefont {Flatt}\ \emph {et~al.}(2023)\citenamefont {Flatt},
  \citenamefont {Busiello}, \citenamefont {Zamuner},\ and\ \citenamefont
  {De~Los~Rios}}]{flatt2023abc}%
  \BibitemOpen
  \bibfield  {author} {\bibinfo {author} {\bibfnamefont {S.}~\bibnamefont
  {Flatt}}, \bibinfo {author} {\bibfnamefont {D.~M.}\ \bibnamefont {Busiello}},
  \bibinfo {author} {\bibfnamefont {S.}~\bibnamefont {Zamuner}}, \ and\
  \bibinfo {author} {\bibfnamefont {P.}~\bibnamefont {De~Los~Rios}},\
  }\href@noop {} {\bibfield  {journal} {\bibinfo  {journal} {Communications
  Physics}\ }\textbf {\bibinfo {volume} {6}},\ \bibinfo {pages} {205} (\bibinfo
  {year} {2023})}\BibitemShut {NoStop}%
\bibitem [{\citenamefont {Liu}\ \emph {et~al.}(2007)\citenamefont {Liu},
  \citenamefont {Guo},\ and\ \citenamefont {Evans}}]{evans}%
  \BibitemOpen
  \bibfield  {author} {\bibinfo {author} {\bibfnamefont {D.-J.}\ \bibnamefont
  {Liu}}, \bibinfo {author} {\bibfnamefont {X.}~\bibnamefont {Guo}}, \ and\
  \bibinfo {author} {\bibfnamefont {J.~W.}\ \bibnamefont {Evans}},\ }\href
  {\doibase 10.1103/PhysRevLett.98.050601} {\bibfield  {journal} {\bibinfo
  {journal} {Phys. Rev. Lett.}\ }\textbf {\bibinfo {volume} {98}},\ \bibinfo
  {pages} {050601} (\bibinfo {year} {2007})}\BibitemShut {NoStop}%
\bibitem [{\citenamefont {Guo}\ \emph {et~al.}(2009)\citenamefont {Guo},
  \citenamefont {Liu},\ and\ \citenamefont {Evans}}]{evans2}%
  \BibitemOpen
  \bibfield  {author} {\bibinfo {author} {\bibfnamefont {X.}~\bibnamefont
  {Guo}}, \bibinfo {author} {\bibfnamefont {D.-J.}\ \bibnamefont {Liu}}, \ and\
  \bibinfo {author} {\bibfnamefont {J.~W.}\ \bibnamefont {Evans}},\ }\href
  {\doibase 10.1063/1.3074308} {\bibfield  {journal} {\bibinfo  {journal} {The
  Journal of Chemical Physics}\ }\textbf {\bibinfo {volume} {130}},\ \bibinfo
  {pages} {074106} (\bibinfo {year} {2009})},\ \Eprint
  {http://arxiv.org/abs/https://pubs.aip.org/aip/jcp/article-pdf/doi/10.1063/1.3074308/15423883/074106\_1\_online.pdf}
  {https://pubs.aip.org/aip/jcp/article-pdf/doi/10.1063/1.3074308/15423883/074106\_1\_online.pdf}
  \BibitemShut {NoStop}%
\bibitem [{\citenamefont {Tom{\'e}}(2006)}]{tome2006}%
  \BibitemOpen
  \bibfield  {author} {\bibinfo {author} {\bibfnamefont {T.}~\bibnamefont
  {Tom{\'e}}},\ }\href@noop {} {\bibfield  {journal} {\bibinfo  {journal}
  {Brazilian journal of physics}\ }\textbf {\bibinfo {volume} {36}},\ \bibinfo
  {pages} {1285} (\bibinfo {year} {2006})}\BibitemShut {NoStop}%
\bibitem [{\citenamefont {Aguilera}\ \emph {et~al.}(2021)\citenamefont
  {Aguilera}, \citenamefont {Moosavi},\ and\ \citenamefont
  {Shimazaki}}]{aguilera2021unifying}%
  \BibitemOpen
  \bibfield  {author} {\bibinfo {author} {\bibfnamefont {M.}~\bibnamefont
  {Aguilera}}, \bibinfo {author} {\bibfnamefont {S.~A.}\ \bibnamefont
  {Moosavi}}, \ and\ \bibinfo {author} {\bibfnamefont {H.}~\bibnamefont
  {Shimazaki}},\ }\href@noop {} {\bibfield  {journal} {\bibinfo  {journal}
  {Nature communications}\ }\textbf {\bibinfo {volume} {12}},\ \bibinfo {pages}
  {1197} (\bibinfo {year} {2021})}\BibitemShut {NoStop}%
\bibitem [{\citenamefont {Pancotti}\ \emph {et~al.}(2020)\citenamefont
  {Pancotti}, \citenamefont {Scandi}, \citenamefont {Mitchison},\ and\
  \citenamefont {Perarnau-Llobet}}]{pancotti2020speed}%
  \BibitemOpen
  \bibfield  {author} {\bibinfo {author} {\bibfnamefont {N.}~\bibnamefont
  {Pancotti}}, \bibinfo {author} {\bibfnamefont {M.}~\bibnamefont {Scandi}},
  \bibinfo {author} {\bibfnamefont {M.~T.}\ \bibnamefont {Mitchison}}, \ and\
  \bibinfo {author} {\bibfnamefont {M.}~\bibnamefont {Perarnau-Llobet}},\
  }\href@noop {} {\bibfield  {journal} {\bibinfo  {journal} {Physical Review
  X}\ }\textbf {\bibinfo {volume} {10}},\ \bibinfo {pages} {031015} (\bibinfo
  {year} {2020})}\BibitemShut {NoStop}%
\bibitem [{\citenamefont {Erdman}\ and\ \citenamefont
  {No{\'e}}(2022)}]{erdman2022driving}%
  \BibitemOpen
  \bibfield  {author} {\bibinfo {author} {\bibfnamefont {P.~A.}\ \bibnamefont
  {Erdman}}\ and\ \bibinfo {author} {\bibfnamefont {F.}~\bibnamefont
  {No{\'e}}},\ }\href@noop {} {\bibfield  {journal} {\bibinfo  {journal} {arXiv
  preprint arXiv:2204.04785}\ } (\bibinfo {year} {2022})}\BibitemShut {NoStop}%
\bibitem [{\citenamefont {Erdman}\ \emph {et~al.}(2023)\citenamefont {Erdman},
  \citenamefont {Rolandi}, \citenamefont {Abiuso}, \citenamefont
  {Perarnau-Llobet},\ and\ \citenamefont {No{\'e}}}]{erdman2023pareto}%
  \BibitemOpen
  \bibfield  {author} {\bibinfo {author} {\bibfnamefont {P.~A.}\ \bibnamefont
  {Erdman}}, \bibinfo {author} {\bibfnamefont {A.}~\bibnamefont {Rolandi}},
  \bibinfo {author} {\bibfnamefont {P.}~\bibnamefont {Abiuso}}, \bibinfo
  {author} {\bibfnamefont {M.}~\bibnamefont {Perarnau-Llobet}}, \ and\ \bibinfo
  {author} {\bibfnamefont {F.}~\bibnamefont {No{\'e}}},\ }\href@noop {}
  {\bibfield  {journal} {\bibinfo  {journal} {Physical Review Research}\
  }\textbf {\bibinfo {volume} {5}},\ \bibinfo {pages} {L022017} (\bibinfo
  {year} {2023})}\BibitemShut {NoStop}%
\bibitem [{\citenamefont {Yeomans}(1992)}]{yeomans1992statistical}%
  \BibitemOpen
  \bibfield  {author} {\bibinfo {author} {\bibfnamefont {J.~M.}\ \bibnamefont
  {Yeomans}},\ }\href@noop {} {\emph {\bibinfo {title} {Statistical mechanics
  of phase transitions}}}\ (\bibinfo  {publisher} {Clarendon Press},\ \bibinfo
  {year} {1992})\BibitemShut {NoStop}%
\bibitem [{\citenamefont {Schnakenberg}(1976)}]{schn}%
  \BibitemOpen
  \bibfield  {author} {\bibinfo {author} {\bibfnamefont {J.}~\bibnamefont
  {Schnakenberg}},\ }\href
  {https://journals.aps.org/rmp/abstract/10.1103/RevModPhys.48.571} {\bibfield
  {journal} {\bibinfo  {journal} {Reviews of Modern physics}\ }\textbf
  {\bibinfo {volume} {48}},\ \bibinfo {pages} {571} (\bibinfo {year}
  {1976})}\BibitemShut {NoStop}%
\bibitem [{\citenamefont {Touchette}(2009)}]{touchette2009large}%
  \BibitemOpen
  \bibfield  {author} {\bibinfo {author} {\bibfnamefont {H.}~\bibnamefont
  {Touchette}},\ }\href@noop {} {\bibfield  {journal} {\bibinfo  {journal}
  {Physics Reports}\ }\textbf {\bibinfo {volume} {478}},\ \bibinfo {pages} {1}
  (\bibinfo {year} {2009})}\BibitemShut {NoStop}%
\bibitem [{\citenamefont {Kumar}\ \emph {et~al.}(2011)\citenamefont {Kumar},
  \citenamefont {Van~den Broeck}, \citenamefont {Esposito},\ and\ \citenamefont
  {Lindenberg}}]{kumar2011thermodynamics}%
  \BibitemOpen
  \bibfield  {author} {\bibinfo {author} {\bibfnamefont {N.}~\bibnamefont
  {Kumar}}, \bibinfo {author} {\bibfnamefont {C.}~\bibnamefont {Van~den
  Broeck}}, \bibinfo {author} {\bibfnamefont {M.}~\bibnamefont {Esposito}}, \
  and\ \bibinfo {author} {\bibfnamefont {K.}~\bibnamefont {Lindenberg}},\
  }\href@noop {} {\bibfield  {journal} {\bibinfo  {journal} {Physical Review
  E}\ }\textbf {\bibinfo {volume} {84}},\ \bibinfo {pages} {051134} (\bibinfo
  {year} {2011})}\BibitemShut {NoStop}%
\bibitem [{\citenamefont {Meibohm}\ and\ \citenamefont
  {Esposito}(2024)}]{Meibohm24}%
  \BibitemOpen
  \bibfield  {author} {\bibinfo {author} {\bibfnamefont {J.}~\bibnamefont
  {Meibohm}}\ and\ \bibinfo {author} {\bibfnamefont {M.}~\bibnamefont
  {Esposito}},\ }\href {\doibase 10.1103/PhysRevE.110.L042102} {\bibfield
  {journal} {\bibinfo  {journal} {Phys. Rev. E}\ }\textbf {\bibinfo {volume}
  {110}},\ \bibinfo {pages} {L042102} (\bibinfo {year} {2024})}\BibitemShut
  {NoStop}%
\bibitem [{\citenamefont {Wachtel}\ \emph {et~al.}(2015)\citenamefont
  {Wachtel}, \citenamefont {Vollmer},\ and\ \citenamefont
  {Altaner}}]{wachtel2015fluctuating}%
  \BibitemOpen
  \bibfield  {author} {\bibinfo {author} {\bibfnamefont {A.}~\bibnamefont
  {Wachtel}}, \bibinfo {author} {\bibfnamefont {J.}~\bibnamefont {Vollmer}}, \
  and\ \bibinfo {author} {\bibfnamefont {B.}~\bibnamefont {Altaner}},\
  }\href@noop {} {\bibfield  {journal} {\bibinfo  {journal} {Physical Review
  E}\ }\textbf {\bibinfo {volume} {92}},\ \bibinfo {pages} {042132} (\bibinfo
  {year} {2015})}\BibitemShut {NoStop}%
\bibitem [{\citenamefont {Gillespie}(1977)}]{gillespie1977exact}%
  \BibitemOpen
  \bibfield  {author} {\bibinfo {author} {\bibfnamefont {D.~T.}\ \bibnamefont
  {Gillespie}},\ }\href@noop {} {\bibfield  {journal} {\bibinfo  {journal} {The
  journal of physical chemistry}\ }\textbf {\bibinfo {volume} {81}},\ \bibinfo
  {pages} {2340} (\bibinfo {year} {1977})}\BibitemShut {NoStop}%
\bibitem [{\citenamefont {Curzon}\ and\ \citenamefont
  {Ahlborn}(1975)}]{curzon1975efficiency}%
  \BibitemOpen
  \bibfield  {author} {\bibinfo {author} {\bibfnamefont {F.}~\bibnamefont
  {Curzon}}\ and\ \bibinfo {author} {\bibfnamefont {B.}~\bibnamefont
  {Ahlborn}},\ }\href@noop {} {\bibfield  {journal} {\bibinfo  {journal}
  {American Journal of Physics}\ }\textbf {\bibinfo {volume} {43}},\ \bibinfo
  {pages} {22} (\bibinfo {year} {1975})}\BibitemShut {NoStop}%
\bibitem [{\citenamefont {Van~den Broeck}(2005)}]{van2005thermodynamic}%
  \BibitemOpen
  \bibfield  {author} {\bibinfo {author} {\bibfnamefont {C.}~\bibnamefont
  {Van~den Broeck}},\ }\href@noop {} {\bibfield  {journal} {\bibinfo  {journal}
  {Physical Review Letters}\ }\textbf {\bibinfo {volume} {95}},\ \bibinfo
  {pages} {190602} (\bibinfo {year} {2005})}\BibitemShut {NoStop}%
\end{thebibliography}%

\newpage
\clearpage

\appendix            

\onecolumngrid
\begin{center}
\textbf{\large Supplemental Material:
{Splitting of nonequilibrium phase transitions in driven Ising models}}
\end{center}
\begin{center}
Gustavo A. L. Forão, Fernando S. Filho, André P. Vieira, Bart Cleuren, Daniel M. Busiello and Carlos E. Fiore
\end{center}

This supplemental material is organized as follows: Sec.~\ref{aptr} introduces key expressions for the all-to-all model, including transition rates and phenomenological descriptions for both ordered phases (collective regimes) discussed in the main text. In Sec.~\ref{apb}, we outline the {leading} coefficients for different thermodynamic quantities near the critical point. The analysis of power {fluctuations} is covered in Sec.~\ref{apc}, while Sec.~\ref{apd} presents additional results on maximum power and efficiency and their relationship to phase transitions.

\section{General all-to-all Ising model and phenomenological descriptions}\label{aptr}

We begin with the all-to-all version of Eq.~(\ref{eqq}), given by
\begin{equation}
E(s) \rightarrow \frac{\epsilon}{2N}\left\{ N_+(N_+-1) + N_-(N_--1) - 2N_+N_- \right\} + \Delta\left(N_+ + N_-\right).
\end{equation}
The energy differences \( \Delta E_{+-} \), \( \Delta E_{+0} \), and \( \Delta E_{-0} \) are defined as
\begin{align}
\Delta E_{+-} &= E_+ - E_- = \frac{2\epsilon}{N}\left(N_+ - N_- + 1\right), \nonumber \\
\Delta E_{+0} &= E_+ - E_0 = \frac{\epsilon}{N}\left(N_+ - N_-\right) + \Delta, \\
\Delta E_{0-} &= E_- - E_0 = \frac{\epsilon}{N}\left(N_+ - N_-\right) - \Delta. \nonumber
\end{align}
In the limit \( N \rightarrow \infty \), where \( N_\pm / N \rightarrow n_\pm \), the expressions for \( \Delta E_{ij} \) and the corresponding transition rates become
\begin{equation}
\omega^{(1)}_{+-} = \Gamma e^{-\frac{\beta_1}{2}\{-2\epsilon(n_- - n_+) + F\}}, \quad
\omega^{(1)}_{+ 0} = \Gamma e^{-\frac{\beta_1}{2}\{\epsilon(n_+ - n_-) + \Delta - F\}}, \quad
\omega^{(1)}_{0-} = \Gamma e^{-\frac{\beta_1}{2}\{\epsilon(n_+ - n_-) - \Delta - F\}}.
\label{a1}
\end{equation}
The reverse transition rates are
\begin{equation}
\omega^{(1)}_{-+} = \Gamma e^{-\frac{\beta_1}{2}\{2\epsilon(n_- - n_+) - F\}}, \quad
\omega^{(1)}_{0+} = \Gamma e^{-\frac{\beta_1}{2}\{-\epsilon(n_+ - n_-) - \Delta + F\}}, \quad
\omega^{(1)}_{-0} = \Gamma e^{-\frac{\beta_1}{2}\{-\epsilon(n_+ - n_-) + \Delta + F\}},
\label{a2}
\end{equation}
with rates \( \omega^{(2)}_{ij} \) obtained from \( \omega^{(1)}_{ij} \) by substituting \( F \rightarrow -F \) and \( \beta_1 \rightarrow \beta_2 \).

As outlined in the main text, steady-state densities are determined by the master equation:
\begin{equation}
{\dot n}_i(t) = \sum_{\nu=1}^2\sum_{j \neq i}\left(\omega^{(\nu)}_{ij}n_j(t) - \omega^{(\nu)}_{ji}n_i(t)\right) = \sum_{\nu=1}^2\sum_{j \neq i} J_{ij}^{(\nu)}(t),
\label{masterall}
\end{equation}
where \( (i,j) \in (0, \pm) \). Since \( n_- + n_0 + n_+ = 1 \), the steady-state densities and {their} associated thermodynamic quantities are derived from two coupled transcendental equations.

To explore {the} thermodynamic behavior {of the system} in strong collective regimes, we apply a phenomenological description {that is valid for} \( |m| \approx 1 \). In this limit, we neglect the density dependence in the transition rates. Approximate steady-state densities are derived using the spanning-tree method \cite{schn}, {following the same approach proposed in \cite{filho2023powerful}. We introduce the following} definitions
\begin{eqnarray}
\gamma_{+} = \tilde{\omega}_{+-}\,\tilde{\omega}_{-0} + \tilde{\omega}_{+0}\,\tilde{\omega}_{0-} + \tilde{\omega}_{+0}\,\tilde{\omega}_{+-} \;, \\
\gamma_0 = \tilde{\omega}_{0-}\,\tilde{\omega}_{-+} + \tilde{\omega}_{0+}\,\tilde{\omega}_{+-} + \tilde{\omega}_{0+}\,\tilde{\omega}_{0-} \;, \\
\gamma_{-} = \tilde{\omega}_{-+}\,\tilde{\omega}_{+0} + \tilde{\omega}_{-0}\,\tilde{\omega}_{0+} + \tilde{\omega}_{-0}\,\tilde{\omega}_{-+},
\end{eqnarray}
where \( \tilde{\omega}_{ij} = \omega^{(1)}_{ij} + \omega^{(2)}_{ij} \).

Using \( n_- = 1 \) and \( n_+ = 0 \) in the transition rates, the steady-state densities for phase \(A\) are
\begin{equation}
n^{\rm st(A)}_{-} = \frac{\gamma_{-}}{\gamma_{+} + \gamma_{0} + \gamma_{-}} \approx \frac{1}{e^{\frac{1}{2} ((\beta_1 + \beta_2) (\Delta + \epsilon) + F (\beta_1 - \beta_2))} + e^{\epsilon (\beta_1 + \beta_2) + \frac{1}{2} F (\beta_2 - \beta_1)} + 1},
\end{equation}
\begin{equation}
n^{\rm st(A)}_{+} = \frac{\gamma_{+}}{\gamma_{+} + \gamma_{0} + \gamma_{-}} \approx \frac{1}{e^{\frac{1}{2} (\beta_1 + \beta_2) (\Delta - \epsilon) + F (\beta_1 - \beta_2)} + e^{\frac{1}{2} F (\beta_1 - \beta_2) - \epsilon (\beta_1 + \beta_2)} + 1},
\end{equation}
and{, as a consequence, we also have:}
\begin{equation}
n^{\rm st(A)}_{0} = \frac{\gamma_{0}}{\gamma_{+} + \gamma_{0} + \gamma_{-}} \approx 1 - n^{\rm st(A)}_{+} - n^{\rm st(A)}_{-}.
\end{equation}

Similarly, for phase \(B\), using \( n_- = 0 \) and \( n_+ = 1 \) in the transition rates, the steady-state densities are
\begin{equation}
n^{\rm st(B)}_{-} = \frac{\gamma_{-}}{\gamma_{+} + \gamma_{0} + \gamma_{-}} \approx \frac{1}{e^{\frac{1}{2} (\beta_1 + \beta_2) (\Delta - \epsilon) - F (\beta_1 - \beta_2)} + e^{\frac{1}{2} F (\beta_2 - \beta_1) - (\beta_1 + \beta_2)\epsilon} + 1},
\end{equation}
\begin{equation}
n^{\rm st(B)}_{+} \approx \frac{\gamma_{+}}{\gamma_{+} + \gamma_{0} + \gamma_{-}} = \frac{1}{e^{\frac{1}{2} ((\beta_1 + \beta_2) (\Delta + \epsilon) - F (\beta_1 - \beta_2))} + e^{(\beta_1 + \beta_2)\epsilon + \frac{1}{2} F (\beta_1 - \beta_2)} + 1},
\end{equation}
{so that:}
\begin{equation}
n^{\rm st(B)}_{0} = \frac{\gamma_{0}}{\gamma_{+} + \gamma_{0} + \gamma_{-}} \approx 1 - n^{\rm st(B)}_{+} - n^{\rm st(B)}_{-}.
\end{equation}

With these approximate steady-state densities, we can calculate all desired thermodynamic quantities, as shown in the main text. Due to their {analytical} complexity, we do not list them here.

\section{Expressions for coefficients in the series expansions of $m$, $\langle \sigma^{(A)} \rangle$ and $\gamma^{(A)}_{\cal P}$ in the case $\Delta=0$}\label{apb}

As stated in the main text, the phase transition at $\epsilon_{1c}$ for $\Delta=0$ is continuous. The time evolution of the order parameter near the criticality can be described by the expansion 
\[
\frac{dm}{dt} \approx a (\epsilon - \epsilon_{1c})m + bm^2 + cm^3 + \dots,
\]
where the coefficient $a$ {has been} defined in the main text as
\[
a = 2 e^{\frac{F}{4} (\beta_1 - \beta_2)} \cosh \left(\frac{F}{4} (\beta_1 - \beta_2)\right) \left(\beta_1 \cosh \left(\frac{\beta_1 F}{2}\right) + \beta_2 \cosh \left(\frac{\beta_2 F}{2}\right)\right),
\]
and $b$ is given by
\[
b = \sinh \left(\frac{F}{4} (\beta_1 - \beta_2)\right) f(\beta_1, \beta_2, F).
\]
{Here,} the function $f(\beta_1, \beta_2, F)$ {can be} explicitly written as {follows}
\begin{align}
f(\beta_1, \beta_2, F) &= -\Bigg[e^{\frac{F}{4} (3 \beta_1 + \beta_2)} 
\bigg(2 \cosh \left(\frac{F}{2} (\beta_1 - \beta_2)\right) + 1\bigg)^2  
\cosh \left(\frac{F}{4} (\beta_1 + \beta_2)\right) \nonumber \\
& \quad \times \Bigg(\left(\beta_1^2 - 4 \beta_1 \beta_2 + \beta_2^2\right) \cosh \left(\frac{F}{4} (\beta_1 - \beta_2)\right) 
- \beta_1^2 \cosh \left(\frac{F}{4} (3 \beta_1 + \beta_2)\right) 
- \beta_2^2 \cosh \left(\frac{F}{4} (\beta_1 + 3 \beta_2)\right)\Bigg) \Bigg] \nonumber \\
& \quad \times \Bigg[\big(e^{\frac{\beta_1 F}{2}} + e^{\frac{\beta_2 F}{2}}\big)^2 
\big(\beta_1 \cosh \left(\frac{\beta_1 F}{2}\right) + \beta_2 \cosh \left(\frac{\beta_2 F}{2}\right)\big)^2 \Bigg]^{-1},
\end{align}
and the coefficient $c$ {reads}
\begin{align}
    c&=\bigg[e^{-\frac{1}{4} F (13 \beta_1+15 \beta_2)} \bigg(e^{F \beta_1}+e^{F \beta_2}+e^{\frac{1}{2} F (\beta_1+\beta_2)}\bigg)^2 \bigg(e^{\frac{1}{2} F (\beta_1+5 \beta_2)} \beta_1^3+e^{5 F \beta_1+2 F \beta_2} \beta_1^3+e^{\frac{1}{2} F (7 \beta_1+\beta_2)} (\beta_1-5 \beta_2) \beta_2 \beta_1+e^{\frac{3}{2} F (3
   \beta_1+\beta_2)}\nonumber\\
   (&5 \beta_1-\beta_2) \beta_2 \beta_1+e^{\frac{7}{2} F (\beta_1+\beta_2)}
   \bigg(\beta_1^2-4 \beta_2 \beta_1-4 \beta_2^2\bigg) \beta_1+e^{\frac{1}{2} F (9 \beta_1+5 \beta_2)}
   \bigg(\beta_1^2+\beta_2^2\bigg) \beta_1+2 e^{2 F (2 \beta_1+\beta_2)} \bigg(\beta_1^2+\beta_2
   \beta_1+2 \beta_2^2\bigg) \beta_1+e^{3 F \beta_1} \beta_2^3+e^{\frac{1}{2} F (5 \beta_1+9 \beta_2)} \beta_2^3\nonumber\\
   &+e^{\frac{3}{2} F (\beta_1+\beta_2)} \beta_2 \bigg(4 \beta_1^2+4 \beta_2 \beta_{1}-\beta_2^2\bigg)+e^{\frac{1}{2} F (5 \beta_1+\beta_2)} \beta_2 \bigg(\beta_1^2+\beta_2^2\bigg)+e^{F (4
   \beta_1+\beta_2)} (\beta_1+\beta_2) \bigg(\beta_1^2-5 \beta_2 \beta_1+\beta_{2}^2\bigg)+e^{\frac{1}{2} F (3 \beta_1+7 \beta_2)} (\beta_1+\beta_2) \bigg(\beta_1^2+\beta_2 \beta_1+\beta_2^2\bigg)\nonumber \\
  +& 2 e^{F (3 \beta_1+\beta_2)} \beta_2 \bigg(2 \beta_1^2+\beta_2 \beta_1+\beta_2^2\bigg)+e^{\frac{1}{2} F (7 \beta_1+3 \beta_2)} (\beta_1+\beta_2) \bigg((\beta_1^2+4 \beta_2 \beta_1+\beta_2^2\bigg)+e^{2 F \beta_1+3 F \beta_2} (\beta_1+\beta_2) \bigg(\beta_1^2+4
   \beta_2 \beta_1+\beta_2^2\bigg)\nonumber\\
   &+e^{3 F \beta_1+2 F \beta_2} (\beta_1+\beta_2) \bigg((\text{$\beta
   $1}^2+7 \beta_2 \beta_1+\beta_2^2\bigg)+3 e^{\frac{5}{2} F (\beta_1+\beta_2)} (\beta_1+\beta_2)
   \bigg((3 \beta_1^2+\beta_2 \beta_1+3 \beta_2^2\bigg)+e^{2 F (\beta_1+2 \beta_2)} \bigg(\beta_1^3-4
   \beta_2 \beta_1^2+2 \beta_2^2 \beta_1-3 \beta_2^3\bigg) \nonumber\\
   &+e^{\frac{1}{2} F (7 \beta_1+5 \beta_2)}
   \bigg((5 \beta_1^3+4 \beta_2 \beta_1^2+6 \beta_2^2 \beta_1-3 \beta_2^3\bigg)+e^{F (\beta_1+2
   \beta_2)} \bigg((2 \beta_1^3+\beta_2 \beta_1^2-\beta_2^3\bigg)+e^{4 F \beta_1+3 F \beta_2} \bigg((2
   \beta_1^3+2 \beta_2 \beta_1^2-\beta_2^3\bigg)\nonumber \\
  & +2 e^{\frac{1}{2} F (3 \beta_1+5 \beta_2)} \bigg((2 \text{$\beta
   $1}^3+2 \beta_2 \beta_1^2+\beta_2^2 \beta_1-\beta_2^3\bigg)+e^{F (\beta_1+3 \beta_2)} \bigg((3
   \beta_1^3-2 \beta_2 \beta_1^2+4 \beta_2^2 \beta_1-\beta_2^3\bigg)+e^{3 F \beta_1+4 F \text{$\beta
   $2}} \bigg(-\beta_1^3+\beta_2^2 \beta_1+2 \beta_2^3\bigg)\nonumber \\
  & +e^{F (2 \beta_1+\beta_2)} \bigg((-\text{$\beta
   $1}^3+2 \beta_2^2 \beta_1+2 \beta_2^3\bigg)+e^{\frac{1}{2} F (5 \beta_1+7 \beta_2)} \bigg((-2 \beta_1^3+2
   \beta_2 \beta_1^2+4 \beta_2^2 \beta_1+4 \beta_2^3\bigg)+e^{2 F (\beta_1+\beta_2)} \bigg((4
   \beta_1^3+13 \beta_2 \beta_1^2+7 \beta_2^2 \beta_1+4 \beta_2^3\bigg)\nonumber \\
   &+e^{3 F (\beta_1+\text{$\beta
   $2})} \bigg(4 \beta_1^3+7 \beta_2 \beta_1^2+13 \beta_2^2 \beta_1+4 \beta_2^3\bigg)+e^{\frac{1}{2} F (5
   \beta_1+3 \beta_2)} \bigg(-3 \beta_1^3+6 \beta_2 \beta_1^2+4 \beta_2^2 \beta_1+5 \beta_2^3\bigg)\bigg) \text{sech}^3\bigg(\frac{1}{4} F (\beta_1-\beta_2)\bigg)\bigg]\nonumber \\
   &\times \bigg[ 128 \bigg((\beta_1 \cosh \bigg(\bigg(\frac{F \text{$\beta
   $1}}{2}\bigg)+\beta_2 \cosh \bigg(\bigg(\frac{F \beta_2}{2}\bigg)\bigg)^3\bigg]^{-1}.
\end{align}

In a similar fashion, the behavior of distinct thermodynamic quantities near the phase transition reinforces the findings in the main text. For instance, the entropy production $\langle \sigma^{(A)} \rangle$ expressed in terms of $m$ near $\epsilon_{1c}$ is given by
\[
\langle \sigma^{(A)} \rangle \sim \langle \sigma_c \rangle + b_\sigma m + c_\sigma m^2 + \dots,
\]
where 
\[
\langle \sigma_c \rangle = 2F\left[\beta_1 \sinh\left(\frac{\beta_1 F}{2}\right) + \beta_2 \sinh\left(\frac{\beta_2 F}{2}\right)\right],
\quad b_\sigma = 0,
\]
and $c_\sigma$ is given by 
\begin{align}
 c_\sigma&=-\bigg[e^{-\frac{3}{2} (\beta _1+\beta _2) F} \bigg(e^{\frac{1}{2} (\beta _1+\beta _2) F}+1\bigg) \bigg(e^{\beta _1 F}+e^{\beta _2
   F}+e^{\frac{1}{2} (\beta _1+\beta _2) F}\bigg) \bigg(-\beta _1^3 F e^{\frac{3 \beta _2 F}{2}}+\beta _1^3 F e^{\bigg(\frac{5 \beta
   _1}{2}+\beta _2\bigg) F}-\beta _2^3 F e^{\frac{3 \beta _1 F}{2}}+\beta _2^3 F e^{\bigg(\beta _1+\frac{5 \beta _2}{2}\bigg) F}\nonumber\\
  & -\bigg(\beta _1-\beta _2\bigg)
   e^{\frac{1}{2} \bigg(4 \beta _1+\beta _2\bigg) F} \bigg(\beta _2^2 F-\beta _2 \bigg(\beta _1 F+4\bigg)+\beta _1 \bigg(\beta _1 F+4\bigg)\bigg)+e^{\frac{3 \beta
   _1 F}{2}+2 \beta _2 F} \bigg(-\bigg(\beta _1^2 \bigg(\beta _1 F+4\bigg)\bigg)+\beta _2 \beta _1 \bigg(\beta _1 F+6\bigg)+2 \beta _2^2 \bigg(\beta _1
   F-1\bigg)\bigg)\nonumber \\
   &+e^{\beta _1 F+\frac{\beta _2 F}{2}} \bigg(\beta _1^2 \bigg(\beta _1 F-4\bigg)+\beta _2 \beta _1 \bigg(6-\beta _1 F\bigg)-2 \beta _2^2
   \bigg(\beta _1 F+1\bigg)\bigg)-\bigg(\beta _1-\beta _2\bigg) e^{\bigg(\beta _1+\frac{3 \beta _2}{2}\bigg) F} \bigg(\beta _1^2 F+2 \beta _1 \bigg(\beta _2
   F+1\bigg)+\beta _2 \bigg(\beta _2 F-2\bigg)\bigg)\nonumber \\
   &-e^{\frac{\beta _1 F}{2}+\beta _2 F} \bigg(2 \beta _1^2 \bigg(\beta _2 F+1\bigg)+\beta _2 \beta _1 \bigg(\beta
   _2 F-6\bigg)+\beta _2^2 \bigg(4-\beta _2 F\bigg)\bigg)+\bigg(\beta _1-\beta _2\bigg) e^{\bigg(\frac{3 \beta _1}{2}+\beta _2\bigg) F} \bigg(\beta _1^2 F+2
   \beta _1 \bigg(\beta _2 F-1\bigg)+\beta _2 \bigg(\beta _2 F+2\bigg)\bigg)\nonumber \\
   &+\bigg(\beta _1-\beta _2\bigg) e^{\frac{1}{2} \bigg(\beta _1+4 \beta _2\bigg) F}
   \bigg(\beta _1^2 F-\beta _1 \bigg(\beta _2 F+4\bigg)+\beta _2 \bigg(\beta _2 F+4\bigg)\bigg)+e^{2 \beta _1 F+\frac{3 \beta _2 F}{2}} \bigg(2 \beta _1^2
   \bigg(\beta _2 F-1\bigg)+\beta _2 \beta _1 \bigg(\beta _2 F+6\bigg)-\beta _2^2 \bigg(\beta _2 F+4\bigg)\bigg)\bigg)\bigg]\nonumber \\
   &\times
   \bigg[ 4 \bigg(e^{\frac{\beta _1
   F}{2}}+e^{\frac{\beta _2 F}{2}}\bigg){}^2 \bigg(\beta _1 \cosh \bigg(\frac{\beta _1 F}{2}\bigg)+\beta _2 \cosh \bigg(\frac{\beta _2 F}{2}\bigg)\bigg){}^2\bigg]^{-1}.
\end{align}

From the scaling behavior \( m \sim (\epsilon_{1c} - \epsilon) \), it follows that \( \langle \sigma^{(A)} \rangle - \langle \sigma_c \rangle \sim (\epsilon_{1c} - \epsilon)^2 \), where \( \delta = 2\beta = 2 \). This result differs from \( \delta = 1 \), which is typically observed in standard order-disorder phase transitions \cite{tome2012, noa2019}.  

The evaluation of \( \langle \dot{Q}_i^{(A)} \rangle \) near \( \epsilon_{1c} \) can be carried out analogously to \( \langle \sigma \rangle \). Specifically, these fluxes behave as \( \langle \dot{Q}_i^{(A)} \rangle \sim \langle \dot{Q}_i^{(c)} \rangle + b_{q_i}m + c_{q_i}m^2 + \dots \), where \( \langle \dot{Q}_i^{(c)} \rangle = -2F \sinh(\beta_i F / 2) \), \( b_{q_i} = 0 \), \( c_{q1} = d_{q1} / d \) and \( c_{q2} = d_{q2} / d \), with
\begin{align}
    d_{q1}=-&\bigg[e^{-F (\frac{3\beta_1}{2}+\beta_2)} (e^{\frac{1}{2} F (\beta_1+\beta_2)}+1) (e^{\frac{1}{2} F (\beta_1+\beta_2)}+e^{\beta_1 F}+e^{\beta_2
   F}) (\beta_1^2 F e^{\beta_2 F}-\beta_1^2 F e^{\frac{1}{2} F (5 \beta_1+\beta_2)}+2 e^{\frac{1}{2} F (\beta_1+\beta_2)} (\beta_1-2 \beta_2+\beta_1\beta_2 F)+ \nonumber\\
   &e^{2\beta_1 F} (\beta_1 (\beta_1 F+4) -2\beta_2
   (\beta_1 F+2))+e^{\frac{3}{2} F (\beta_1+\beta_2)} (\beta_1 (\beta_1
   F+4)-\beta_2 (\beta_1 F+2))-2 e^{F (2\beta_1+\beta_2)} (2\beta_2+\beta_1 (\beta_2 F-1))+ \nonumber\\
   &e^{\frac{1}{2} F (\beta_1+3\beta_2)} (\beta_1 (\beta_1
   (-F)+2\beta_2 F+4)-4\beta_2)+e^{\beta_1 F} (\beta_1 (F (\beta_2-\beta_1)+4)-2\beta_2)-e^{\frac{1}{2} F (3\beta_1+\beta_2)} (2\beta_2+\beta_1 (F
   (\beta_1+\beta_2)-2))+ \nonumber \\ 
   &e^{F (\beta_1+\beta_2)} (\beta_1 (F 
   (\beta_1+\beta_2)+2)-2\beta_2)) \bigg],
\end{align}
\begin{align}
  d_{q2}= & e^{-F (\beta_1+\frac{3 \beta_2}{2})} (e^{\frac{1}{2} F (\beta_1+\beta_2)}+1) (e^{\frac{1}{2} F (\beta_1+\beta_2)}+e^{\beta_1 F}+e^{\beta_2
   F}) (\beta_2^2 F \left(-e^{\beta_1 F}\right)+\beta_2^2 F e^{\frac{1}{2} F (\beta_1+5 \beta_2)}+e^{\frac{1}{2} F (3 \beta_1+\beta_2)} (\beta_1 (4-2 \beta_2
   F)+\beta_2 (\beta_2 F-4))\nonumber \\
   &+e^{\beta_2 F} (\beta_1 (2-\beta_2 F)+\beta_2
   (\beta_2 F-4))-2 e^{\frac{1}{2} F (\beta_1+\beta_2)} (\beta_2+\beta_1
   (\beta_2 F-2))+2 e^{F (\beta_1+2 \beta_2)} (\beta_1 (\beta_2 F+2)-\beta_2)\nonumber \\
   &+e^{\frac{1}{2} F (\beta_1+3 \beta_2)} (\beta_1 (\beta_2 F+2)+\beta_2
   (\beta_2 F-2))-e^{F (\beta_1+\beta_2)} (\beta_1 (\beta_2 F-2)+\beta_2
   (\beta_2 F+2))+e^{\frac{3}{2} F (\beta_1+\beta_2)} (\beta_1 (\beta_2
   F+2)-\beta_2 (\beta_2 F+4))+ \nonumber \\
   &e^{2 \beta_2 F} (2 \beta_1 (\beta_2
   F+2)-\beta_2 (\beta_2 F+4))),
\end{align}
and
\[
d = 4\left(e^{\frac{\beta_1 F}{2}} + e^{\frac{\beta_2 F}{2}}\right)^2\left(\beta_1 \cosh\left(\frac{\beta_1 F}{2}\right) + \beta_2 \cosh\left(\frac{\beta_2 F}{2}\right)\right)^2.
\]

As a {consequence}, the power \( \langle \mathcal{P}^{(A)} \rangle = -\langle \dot{Q}_1^{(A)} \rangle - \langle \dot{Q}_2^{(A)} \rangle \) and the efficiency \( \eta^{(A)} = 1 + \langle \dot{Q}_1^{(A)} \rangle / \langle \dot{Q}_2^{(A)} \rangle \) exhibit the same scaling behavior. We also note that the coefficient \( c_\sigma \) is related to \( c_{q1} \) and \( c_{q2} \) through the relation \( c_\sigma = -\beta_1 c_{q1} - \beta_2 c_{q2} \).  

Conversely, the bistability observed at \( \epsilon_{2c} \) is also reflected in all thermodynamic quantities. However, the linear analysis is unsuitable in this case. Instead, numerical solutions of the full master equations for \(m\) and \(q\) can be substituted into the expressions for thermodynamic quantities, as illustrated in Fig.~(\ref{fluc}) in the main text.

\section{Evaluation of the power variance $\gamma_{\cal P}$}\label{apc}

In this section, we {evaluate power fluctuations that are captured by the power variance}, \( \gamma_{\mathcal{P}} \).  

For the all-to-all topology, the evaluation employs the large deviation method \cite{touchette2009large, kumar2011thermodynamics}. Let \( P(i, \mathcal{P}, t) \) denote the probability of the system being in state \( i \) at time \( t \) with power \( \mathcal{P} \). The time evolution of \( P(i, \mathcal{P}, t) \) is governed by the master equation 
\[
\frac{\partial}{\partial t} P(i, \mathcal{P}, t) = \sum_{\nu=1}^2 \sum_{j} \left\{ \omega_{ij}^{(\nu)} P(j, \mathcal{P} - \Delta \mathcal{P}, t) - \omega_{ji}^{(\nu)} P(i, \mathcal{P}, t) \right\},
\]
where \( \omega_{ji}^{(\nu)} \) is the transition rate as defined in Sec.~\ref{aptr}, and \( \Delta \mathcal{P} = \pm F \) for transitions between states \( j \) and \( i \), {where different signs are associated with} clockwise or counterclockwise transitions with the system in contact with the cold and hot thermal baths, respectively (see the main text {for modeling details}).

The characteristic function of the power is defined as \( \rho_p(i, \alpha, t) = \int_{-\infty}^{\infty} d\mathcal{P} \, e^{-\alpha \mathcal{P}} P(i, \mathcal{P}, t) \), whose time evolution is described by  
\[
\frac{\partial}{\partial t} \rho_p(i, \alpha, t) = \sum_{\nu=1}^2 \sum_{j} \left\{ \omega_{ij}^{(\nu)} \rho_p(j, \alpha, t) e^{-\alpha \Delta \mathcal{P}} - \omega_{ji}^{(\nu)} \rho_p(i, \alpha, t) \right\}.
\]
This equation can be rewritten as  
\[
\frac{\partial \rho_p(i, \alpha, t)}{\partial t} = M_p(\alpha) \rho_p(i, \alpha, t),
\]  
where \( M_p(\alpha) \) is the tilted matrix, explicitly given as
\[
M_p(\alpha) = 
\begin{pmatrix}
-(\omega_{0-}^{(1)} + \omega_{+-}^{(1)} + \omega_{0-}^{(2)} + \omega_{+-}^{(2)}) & \omega_{-0}^{(1)} e^{-\alpha F} + \omega_{-0}^{(2)} e^{\alpha F} & \omega_{-+}^{(1)} e^{\alpha F} + \omega_{-+}^{(2)} e^{-\alpha F} \\
\omega_{0-}^{(1)} e^{-\alpha F} + \omega_{0-}^{(2)} e^{\alpha F} & -(\omega_{-0}^{(1)} + \omega_{+0}^{(1)} + \omega_{-0}^{(2)} + \omega_{+0}^{(2)}) & \omega_{0+}^{(1)} e^{-\alpha F} + \omega_{0+}^{(2)} e^{\alpha F} \\
\omega_{+-}^{(1)} e^{-\alpha F} + \omega_{+-}^{(2)} e^{\alpha F} & \omega_{+0}^{(1)} e^{\alpha F} + \omega_{+0}^{(2)} e^{-\alpha F} & -(\omega_{-+}^{(1)} + \omega_{0+}^{(1)} + \omega_{-+}^{(2)} + \omega_{0+}^{(2)})
\end{pmatrix}.
\]

Using the large-deviation method \cite{touchette2009large, kumar2011thermodynamics}, the scaled cumulant generating function is determined by the largest eigenvalue \( \lambda_p(\alpha) \) of \( M_p(\alpha) \). From \( \lambda_p(\alpha) \), the mean \( \langle \mathcal{P} \rangle \) and variance \( \gamma_{\mathcal{P}} \) are given by  

\[
\langle \mathcal{P} \rangle = \left. \frac{\partial \lambda_p(\alpha)}{\partial \alpha} \right|_{\alpha=0}, \quad \gamma_{\mathcal{P}} = \left. \frac{\partial^2 \lambda_p(\alpha)}{\partial \alpha^2} \right|_{\alpha=0}.
\]

Since the expression for \( \lambda_p(\alpha) \) is generally cumbersome, we instead use an alternative approach \cite{wachtel2015fluctuating}. In this method, the coefficients of the characteristic function satisfy the cubic equation  
\[
a_0 + a_1 \lambda_p(\alpha) + a_2 \lambda_p^2(\alpha) - \lambda_p^3(\alpha) = 0,
\]  
allowing \( \langle \mathcal{P} \rangle \) and \( \gamma_{\mathcal{P}} \) to be evaluated as  
\[
\langle \mathcal{P} \rangle = -\frac{1}{a_1} \left( \frac{\partial a_0}{\partial \alpha} \right)_{\alpha=0}, \quad \gamma_{\mathcal{P}} = \frac{1}{a_1^3} \left[ a_1^2 \frac{\partial^2 a_0}{\partial \alpha^2} - 2a_1 \frac{\partial a_0}{\partial \alpha} \frac{\partial a_1}{\partial \alpha} + 2a_2 \left( \frac{\partial a_0}{\partial \alpha} \right)^2 \right]_{\alpha=0}.
\]  

Exact values of \( \langle \mathcal{P} \rangle \) and \( \gamma_{\mathcal{P}} \) are obtained by inserting the steady-state probabilities \( \{p_i^{\text{st}}\} \) into these derivatives. Approximate expressions for \( \gamma_{\mathcal{P}}^{(A)} \) and \( \gamma_{\mathcal{P}}^{(B)} \) in the phenomenological description are obtained by assuming \( n_- = 1 \) (\( n_+ = 0 \)) and \( n_- = 0 \) (\( n_+ = 1 \)), respectively, in the transition rates.  

At \( \epsilon_{1c} \), \( \gamma_{\mathcal{P}}^{(A)} \) behaves as \( \gamma_{\mathcal{P}}^{(A)} \sim (\gamma_{\mathcal{P}})^{(c)} + b_v m + c_v m^2 + \ldots \), where \( b_v = 0 \), and the expression for \( c_v \) is omitted due to its length.  

For the square-lattice topology, \( \gamma_{\mathcal{P}} \) is computed via Gillespie simulations \cite{gillespie1977exact}. For the \( i \)-th stochastic trajectory up to time \( t_{\text{max}} \), the total power {and its square are} recorded as \( \mathcal{P}_i = \sum_{t=0}^{t_{\text{max}}} \mathcal{P}_{it} \) and \( \mathcal{P}_i^2 = \sum_{t=0}^{t_{\text{max}}} \mathcal{P}_{it}^2 \). Repeating this process \( M \) times (with \( M = 10^4 \)), the averages are calculated as  
\[
\langle \mathcal{P} \rangle_{t_{\text{max}}} = \frac{1}{M} \sum_{i=1}^M \mathcal{P}_i, \quad \langle \mathcal{P}^2 \rangle_{t_{\text{max}}} = \frac{1}{M} \sum_{i=1}^M \mathcal{P}_i^2.
\]
Finally, \( \langle \mathcal{P} \rangle = \langle \mathcal{P} \rangle_{t_{\text{max}}} / t_{\text{max}} \) and \( \gamma_{\mathcal{P}} = (\langle \mathcal{P}^2 \rangle_{t_{\text{max}}} - \langle \mathcal{P} \rangle_{t_{\text{max}}}^2) / t_{\text{max}} \).  

Figure~\ref{fluc} compares \( \gamma_{\mathcal{P}} \) for all-to-all (continuous and dashed lines) and square-lattice topologies (symbols \( \times \) for a system size \( N = 10^2 \)). As shown, the estimates agree well.


\begin{figure}[h!]
\centering
    \includegraphics[width=0.6 \textwidth]{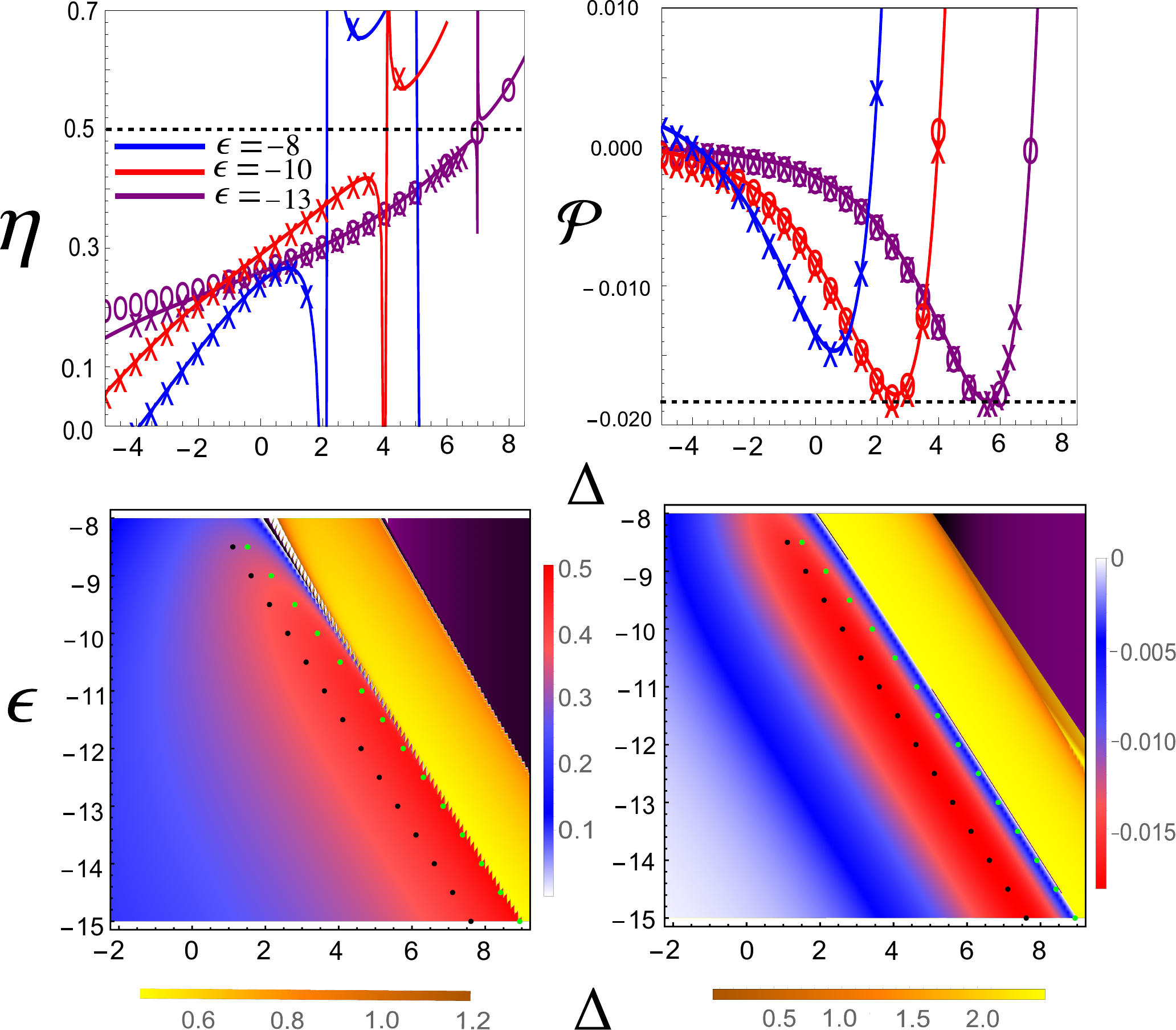}
    \caption{
    Upper panels: Plots of efficiency \( \eta \) (left) and power \( \mathcal{P} \) (right) as functions of \( \Delta \) for different values of \( \epsilon \), with driving \( F = 2 \). Solid lines represent all-to-all mean-field results, \(\times\) symbols denote square lattice results, and circles show phenomenological predictions from Eqs.~(\ref{prodef})-(\ref{poweref}). Dashed lines mark ideal efficiency \( \eta_{c} = 1 - \beta_2 / \beta_1 \) (left) and maximum power \( \mathcal{P}_{mP} = -0.01831 \) (right). Lower panels: Heat maps of efficiency and power. Black and green dots indicate MP and ME points, respectively. The transition from heat engine (red) to pump (yellow) approaches ideal efficiency \( \eta = \eta_c \). Parameters: \( \beta_1 = 2 \), \( \beta_2 = 1 \), and \( F = 2 \).}
    \label{paneld}
\end{figure}

\section{Extra results for heat engines operating on the verge of maximum power and maximum efficiency}\label{apd}

As discussed in the main text, a novel feature of our system is the existence of a small region where the heat engine operates close to maximum power and efficiency. Figures~\ref{fig3} and \ref{paneld} illustrate the maximization of power and efficiency with respect to $\Delta$. These plots show how $\Delta$ varies for a fixed $\beta_2 = 1$ and explore the behavior of the system under different values of $\epsilon$ and $F$. 

Three remarkable aspects are worth highlighting: 
\begin{enumerate}
\item The efficiencies at maximum power, $\eta_{MP}$, are constrained between the Curzon-Ahlborn efficiency, $\eta_{CA} = 1 - \sqrt{\beta_2/\beta_1}$, and the maximum efficiency, $\eta_{ME}$. Although $\eta_{CA}$ is not a universal bound, it depends solely on the temperature ratio and has been verified for various heat engines operating out of equilibrium \cite{curzon1975efficiency,van2005thermodynamic}.

\item As $\beta_1$ increases, both $\eta_{MP}$ and $\eta_{ME}$ converge, eventually matching the ideal Carnot efficiency $\eta_c = 1 - \beta_2/\beta_1$ as $\epsilon$ and $F$ increase. However, this optimization of power and efficiency comes at the cost of increased dissipation, as indicated by the rise in ${\cal P}_{MP}$ (see the inset).

\item The points of maximum power, $\Delta_{MP}$, and maximum efficiency, $\Delta_{ME}$, approach the discontinuous phase transition point $\Delta_{1c}$ as $\beta_1$ increases. This reveals a striking relationship between optimization and the occurrence of a phase transition.
\end{enumerate}

\end{document}